\documentclass[aps,tightenlines,groupedaddress,superscriptaddress,
preprint,
nofootinbib]{revtex4}

\usepackage{graphicx}
\usepackage{amsmath}
\usepackage{bm}
\usepackage{colordvi}
\usepackage{color}
\usepackage{epsfig}
\usepackage{fancybox}
\usepackage{rotating}

\usepackage{relsize}
\def\babar{\mbox{\slshape B\kern-0.1em{\smaller A}\kern-0.1em
    B\kern-0.1em{\smaller A\kern-0.2em R}}}
\newcommand{\DZero}{\rm D\O}

\begin{document}


\title{Quarkonium at the Frontiers of High Energy Physics:\\
A Snowmass White Paper}


\author{Geoffrey T.~Bodwin}
\affiliation{HEP Division, Argonne National Laboratory,
9700 South Cass Avenue, Lemont, IL 60439, USA}
\author{Eric Braaten}
\affiliation{Department of Physics, The Ohio State University, 
Columbus, OH 43210, USA}
\author{Estia Eichten}
\affiliation{Theoretical Physics Department, Fermilab, IL 60510, 
USA}
\author{Stephen Lars Olsen}
\affiliation{Department of Physics and Astronomy, Seoul National University, 
Seoul 151-747, Korea}
\author{Todd K.~Pedlar}
\affiliation{Department of Physics, Luther College, 
Decorah, IA 52101, USA}
\author{James Russ}
\affiliation{Carnegie-Mellon University, Pittsburgh, PA 15213, USA}

\date{\today}

\begin{abstract}
In this Snowmass White Paper, we discuss physics opportunities
involving heavy quarkonia at the intensity and energy frontiers of high
energy physics. We focus primarily on two specific aspects of quarkonium
physics for which significant advances can be expected from experiments
at both frontiers. The first aspect is the spectroscopy of charmonium and
bottomonium states above the open-heavy-flavor thresholds. Experiments at
$e^+ e^-$ colliders and at hadron colliders have discovered many new,
unexpected quarkonium states in the last 10 years. Many of these states
are surprisingly narrow, and some have electric charge. The observations
of these charged quarkonium states are the first definitive discoveries
of manifestly exotic hadrons. These results challenge our understanding
of the QCD spectrum. The second aspect is the production of heavy
quarkonium states with large transverse momentum. Experiments at the LHC
are measuring quarkonium production with high statistics at
unprecedented values of $p_T$. Recent theoretical developments may
provide a rigorous theoretical framework for inclusive production of
quarkonia at large $p_T$. Experiments at the energy frontier will
provide definitive tests of this framework. Experiments at the intensity
frontier also provide an opportunity to understand the exclusive
production of quarkonium states.
\end{abstract}

\maketitle


%

\section{Executive Summary}

\subsection{Introduction}

The discovery of charmonium in the November revolution of 1974 was a
watershed moment in the establishment of Quantum Chromodynamics (QCD) as
the fundamental field theory of the strong force and, more generally, in
the establishment of the Standard Model of particle physics. Since that
time, theoretical and experimental investigations in quarkonium physics
have continued to be pillars of the international research effort in QCD,
and the source of many new insights.

There are several reasons for this continuing interest and activity in
quarkonium physics. One reason is that the nonrelativistic nature of
heavy-quarkonium systems makes it easier to unravel the complicated
effects of QCD dynamics. This allows one to use powerful
effective-field-theory tools in the theoretical analyses of these
systems. Another reason is that some quarkonium systems have
particularly clean experimental signatures, which allows detailed
studies with high statistics even in the challenging environment of a
hadron collider at the energy frontier. Finally, quarkonium systems
provide a unique laboratory in which to understand the interplay between
perturbative and nonperturbative effects in QCD, because the heavy-quark
mass provides a natural boundary between the perturbative and
nonperturbative regimes.

In this Snowmass White Paper, we describe two aspects of quarkonium
physics in which dramatic progress can be expected through an interplay
of theory with experiments at the frontiers in high-energy physics: (1)
the spectroscopy of quarkonia above the open-heavy-flavor thresholds,
and (2) the production of quarkonia at large transverse momentum.
Progress on both of these problems will be driven by experiments at both
the energy frontier and the intensity frontier. In addition to the
dramatic progress that can be expected in these two areas, incremental
progress can be expected in many other areas of quarkonium physics. See
Refs.~\cite{Brambilla:2004wf,Brambilla:2010cs} for detailed accounts of
the opportunities in these other areas.

\subsection{Activity in Quarkonium Physics \label{sec:QWG}}

The strong, continuing interest in quarkonium physics within the high
energy physics community is well illustrated by the activities of the
Quarkonium Working Group (QWG). The aim of this organization, which
consists of about 100 high-energy physicists, is to further interactions
between experiment and theory on quarkonium physics, to identify
problems at the forefront of quarkonium physics, and to facilitate the
solutions of those problems. The QWG has sponsored the writing of two
extensive reviews on quarkonium physics in 2004 and in 2010, which have
highlighted the frontier issues in theory and experiment
\cite{Brambilla:2004wf,Brambilla:2010cs}. The impacts of these
reviews are evident in their numbers of citations, which have
reached 600 and 400 as of August 2013, respectively. The QWG sponsors
an international workshop on quarkonium physics on a regular basis.  The
most recent QWG workshop was held in Beijing in April 2013, and the next
one will be held at CERN in November 2014. Other physics groups also
sponsor conferences that are devoted to quarkonium physics, with
several such conferences typically taking place each year.

The activity in quarkonium physics at the intensity frontier is
illustrated dramatically by the impact of the $B$-factory experiments in
this area. The most highly cited paper by the Belle collaboration is its
2003 paper on the discovery of the $X(3872)$, which has over 800
citations as of August 2013. Of Belle's 15 most highly cited papers, 5
are on quarkonium physics. The $4^{\rm th}$ most highly cited paper by
the BaBar collaboration is their 2005 paper on the discovery of the
$Y(4260)$, which has about 470 citations as of August 2013. Of Babar's
30 most highly cited papers, 9 are on quarkonium physics. Given that
quarkonium physics was barely  mentioned as a motivation for these
experiments, their impacts in this area have been especially remarkable.
There is every reason to expect that the intensity-frontier experiments,
such as Belle~II, and the energy-frontier experiments, such as LHCb,
will have a comparable impacts on quarkonium spectroscopy.

One indication of the activity in quarkonium physics at the energy
frontier is the number of quarkonium related papers that have emerged
from the experimental collaborations at the LHC. To date, about 250
LHC quarkonium papers have been submitted to the arXiv. 
While the bulk of these results have been in the areas of production and
spectroscopy, some also address other areas of quarkonium physics. This
high level of activity can be expected to continue at future runs of the
LHC.

\subsection{Quarkonium Spectroscopy}

The spectroscopy of heavy quarkonia above the open-heavy-flavor
threshold is an aspect of quarkonium physics in which dramatic progress
can be expected through experiments at the intensity frontier
and theoretical efforts to understand the results. 
Experiments have uncovered entirely new exotic hadronic spectroscopies                                                           
in the open-charm and open-bottom threshold regions.
Understanding the QCD spectrum in these regions
is one of the most interesting and exciting challenges to theorists 
today.
Quarkonium systems provide a relatively clean environment
to probe strong dynamics in the threshold regions.
The lessons learned
should be more generally applicable
to the more complicated situation with the light hadron spectrum.

One of the basic properties of QCD is 
the hadron spectrum.
The elementary constituents of hadrons are quarks ($q$), antiquarks ($\bar q$),
and gluons ($g$). Their interactions are described by QCD, 
a vector $SU(3)$ color force between quarks ($3$) and antiquarks ($\bar 3$)
mediated by an octet of gluons. There are six flavors of quarks  
($u,d,s,c,b,t$) observed in nature. The masses for the $u$, $d$, and $s$ quarks 
are small compared to the scale of QCD ($\Lambda_{\rm QCD}$) 
while the $c$, $b$, and $t$ quarks are heavy.  
QCD dynamics simplify in both the $m_{\rm quark} \rightarrow 0$ limit, 
because of the resulting chiral symmetries, 
and in the $m_{\rm quark} \rightarrow \infty$  limit,
because of heavy quark symmetry and the possibility of using 
nonrelativistic effective field theories to describe the heavy quarks.

Because QCD is confining, only color-singlet hadron states exist in nature.  
The low-lying spectrum of hadrons are usually
classified by their valence (flavor) quark content. 
In the naive quark model, mesons are $q\bar q$ states 
and baryons are $qqq$ states.  
For heavy quarks, the low-lying mesons are called charmonium  ($c\bar c$) 
and bottomonium ($b \bar b$)  or, collectively, quarkonium.  
Below the open-heavy-flavor thresholds for pairs of heavy-light mesons 
($D\bar D$ and $B \bar B$, respectively),  
these quarkonium states are narrow
and are, thus, ideal systems for precision studies of QCD dynamics.    
Low-lying $b\bar c$ systems also have these properties.
Hadronic states involving $t$ quarks do not have a chance to form 
before the $t$ quark decays through weak interactions. 

Because quarkonium systems contain heavy quarks, one expects 
nonrelativistic dynamics to apply. 
Since the discovery of the first charmonium state, the $J/\psi$, 
in 1974, simple models that treat the complicated QCD dynamics 
of the gluons and light quarks as an effective potential 
between the heavy quark-antiquark have been employed. 
This dynamical assumption is the Born-Oppenheimer approximation  
commonly used in atomic and molecular physics.  
Potential models enjoyed tremendous success in the early years --  
accurately predicting the excitation spectrum and 
the electromagnetic and 
hadronic transitions among these narrow quarkonium states.  

With the advent in the late 90's of a new generation of 
high-luminosity $B$ factories (PEPII, KEKB, and CESRII)
and a high-luminosity charm factory (BEPCII),  more detailed studies 
of the properties of the known states were possible as well as  
the observation of many previously unobserved narrow spin-singlet states  
in the charmonium spectrum ($\eta_c$, $\eta_c^{\prime}$, $h_c$) 
and in the bottomonium spectrum 
($\eta_b$, $\eta_b'$, $h_b$, $h_b'$, \ldots). 
Two contemporaneous theoretical developments  
put the studies of quarkonium on more rigorous grounds.  
Effective field theories (NRQCD,  potential NRQCD, ...) 
enabled rigorous calculations using perturbation theory.
Lattice QCD simulations with full QCD dynamics,
except for the $b$ quark, enabled calculations 
of masses and decays of the low-lying states 
of quarkonium systems.
All these results strengthened the agreement 
with theory expectations,  
at least for quarkonia below the open-heavy-flavor thresholds.

A decade ago, a new state, the $X(3872)$, was observed at Belle
and quickly confirmed by CDF, \DZero, and $\babar$.  Initial attempts 
to identify it as a missing state of the conventional charmonium 
spectrum failed as more of its properties were measured.  
The $X(3872)$ mass
is very close to a strong decay threshold ($D^{*0} \bar D^0$) 
and its decays exhibit
significant isospin violation.  This was a 
surprise and upset the simple picture 
that had been so successful for 30 years.  
New states and new surprises followed quickly.  
Additional states have been discovered in the threshold regions 
for both the charmonium and
bottomonium systems.  They have been collectively denoted as the $XYZ$ states.  
Some of these new states have conventional quarkonium interpretations, 
but others clearly do not fit.  Particularly difficult to interpret is  
the $Y(4260)$ state:  a $1^{--}$ state that is 
only barely observable as an 
$s$-channel resonance in $e^+e^-$ collisions and 
which appears at an energy where no conventional charmonium state 
is expected. 
Finally, a number of charged quarkonium states have now been observed.
The most well-established are the charged bottomonium states   
$Z^+_b(10610)$ and $Z^+_b(10650)$ and the charged charmonium state
$Z^+_c(3900)$.  Clearly such states must have tetraquark 
structures. The present list of $XYZ$ states
extends to almost two dozen states.  

In retrospect, the failure of the model of mesons as 
simple $q\bar q$ states is to be expected, because QCD has a much
richer structure than the quark model.  The excitation spectra 
of quarkonium systems reflect these additional degrees of freedom.  
Exciting the gluonic (string) degrees of freedom 
is expected to produce hybrid states.  
Tetraquark states with additional light-quark valence pairs 
states ($Q\bar q\bar Q q'$) should appear near the 
open-heavy-flavor thresholds.  
The dynamics of such tetraquark  states 
has many possibilities (meson molecules, compact tetraquarks, 
diquarkonium, hadroquarkonium, \ldots).   Lattice QCD groups have made 
preliminary calculations of the spectrum of charmonium that reveal 
the low-lying hybrid states as well as the conventional charmonium
states.   Lattice calculations of tetraquark systems are also possible, 
but these are technically very difficult at the present time.  

The next generation of collider experiments 
with capabilities of studying quarkonium states 
(BESIII, LHCb, Belle~II, PANDA) are online now
or will be in the near future.  With much higher production rates 
for quarkonium states, we can expect valuable clues to help 
disentangle  the physics of the open-heavy-flavor threshold region. 
Hadron collider experiments  (ATLAS, CMS, LHCb)  will likely provide
the first detailed studies of the $b\bar c$ system as well.

\subsection{Quarkonium Production}

The production of heavy quarkonia at large transverse momentum $p_T$ is
an aspect of quarkonium physics in which dramatic progress can be
expected through the interaction of theory with experiments at the
energy frontier. The best prospects for an understanding of quarkonium
production that is based rigorously on QCD are in the region of
$p_T$ that is much larger than the quarkonium mass.  The Tevatron
experiments began to reach this region in their measurements of
charmonium production, and the LHC experiments have begun to reach
this region in their measurements of bottomonium production. In future
runs of the LHC, both charmonium and bottomonium production will be
measured deep into the relevant large-$p_T$ regions, with very
high statistics.  This will provide powerful tests of theoretical
frameworks for quarkonium production.

Interactions between theory and experiment have played a key role in the
history of quarkonium production. Since the first measurements of
quarkonium production cross sections in CDF Run~I, quarkonium production
experiments have stimulated theoretical developments, which have, in
turn, motivated increasingly extensive and challenging experimental
measurements. During the last two decades, this process has continued at
an intense level and has led to many impressive achievements in both
experiment and theory.

Quarkonium production continues to be relevant to the international
high-energy physics program in numerous ways. Quarkonium production is
interesting in its own right, as an aspect of QCD that is not fully
understood. It is also useful as a laboratory for exploring experimental
techniques within a setting in which experimental signatures are very
clean and very high statistics can be accumulated. Quarkonium production
is also useful as a theoretical laboratory for QCD.   It can be used to
test and extend our understanding of factorization theorems, which are
the theoretical foundation for all perturbative calculations in QCD. New
theoretical concepts that have been developed in order to understand
large higher-order perturbative corrections to quarkonium production,
such as those that arise from kinematic enhancements and from large
endpoint logarithms, could have wider applicability in the calculation
of high-energy cross sections. Insights gained from studying quarkonium
production are also relevant to physics beyond the Standard Model. For
example, certain quarkonium production processes can be used to measure
Higgs couplings and so to probe for physics beyond the standard model.
If that new physics involves nonrelativistic bound states, then the
techniques that have been developed for understanding quarkonium
production will be directly applicable.

The current standard method for calculating quarkonium production rates
is the nonrelativistic QCD (NRQCD) factorization approach, which is
based on the effective field theory NRQCD. In this approach, production
rates are expressed as perturbatively calculable partonic cross sections
multiplied by nonperturbative constants called NRQCD matrix elements.
Some of the NRQCD matrix elements must be determined through fits of
NRQCD factorization predictions to experimental data, but they can then
be used to make predictions for other quarkonium production
processes. Universality of the NRQCD matrix elements is an essential
feature of NRQCD factorization, and so it is important to test the
predictions of NRQCD factorization in as many processes as possible. The
NRQCD factorization approach is a conjecture that has not been proven to
all orders in $\alpha_s$. Therefore, at present, it must be regarded as
a model, to be tested by experiment, rather than as a consequence of
QCD. Since standard methods for proving factorization fail unless
the hard-scattering momentum transfer $p_T$ is much larger than the
initial- and final-state hadron masses, it seems likely that NRQCD
factorization, if it is correct, holds only for $p_T$ much larger than
the quarkonium mass.

An important recent theoretical development is the next-to-leading-power
(NLP) fragmentation approach, in which quarkonium production rates are
expressed as perturbatively calculable partonic cross sections convolved
with fragmentation functions, up to corrections suppressed by a
factor $m_Q^4/p_T^4$, where $m_Q$ is the heavy-quark mass. Unlike
the NRQCD factorization formula, the NLP factorization formula has
been proven to all orders in $\alpha_s$. The NLP fragmentation approach
becomes more predictive if NRQCD factorization is used to
express the fragmentation functions in terms of NRQCD matrix elements.
The NLP fragmentation approach then organizes the NRQCD
factorization expression for the cross section according to powers
of $m_Q^2/p_T^2$, which facilitates the calculation of higher-order
corrections and the resummation of large logarithms.

NRQCD factorization predictions have now been computed at
next-to-leading order (NLO) in $\alpha_s$ for many production
processes. In general, these predictions agree with the experimental
data from $pp$, $p\bar p$, $ep$, and $e^+e^-$ colliders for the
production of $S$-wave and $P$-wave quarkonia at large $p_T$. The most
notable exception is the polarization of the $J/\psi$, $\psi(2S)$, and
$\Upsilon(nS)$ ($n=1,2,3$) at the Tevatron and at the LHC. This
serious discrepancy between theory and experiment deserves further
investigation from both the theoretical and experimental sides.
Definitive measurements of the polarizations will be carried out at
the LHC. An important issue with regard to quarkonium polarizations is
that prompt $J/\psi$ and $\Upsilon(1S)$ production rates include feeddown
from $S$-wave and $P$-wave quarkonium states of higher mass. The
inclusion of feeddown in the theoretical predictions brings in
additional theoretical uncertainties. Therefore, it would be very useful
for future experiments to separate the feeddown contributions from
direct-production contributions.

At the frontiers of high-energy physics, there are many opportunities
for further interesting work on quarkonium production.
In Run~2 of the LHC, the extension of the energy frontier to $13$~TeV
and the increase in luminosity in comparison with Run~I will make it
possible to extend the $p_T$ reach of quarkonium studies. Since NRQCD
factorization, if it is correct, is expected to hold only for values of
$p_T$ that are much larger than the quarkonium mass, it is important to
measure quarkonium production with high statistics at the highest
possible values of $p_T$. Measurements of production and polarization
for the $\chi_{cJ}$ and $\chi_{bJ}$ states and for new processes, such
as associated production with $W$ or $Z$ bosons, would provide valuable
additional tests of the theory. $B_c$ studies could also provide new
insights into the production of mesons with nonzero flavor.

While many of the key tests of the theory of charmonium production 
could be accomplished in upcoming runs of the LHC, the LHC
upgrade would afford the opportunity to push studies of $b\bar b$ states
to still higher values of $p_T$. The $b\bar b$ systems are particularly
important tests of the validity of NRQCD because they are more
nonrelativistic than the $c\bar c$ systems. Precision measurements of
cross sections and polarizations of $b\bar b$ states at values of $p_T$
that are much larger than the quarkonium mass will be challenging, but
they are crucial tests of the NRQCD factorization approach.

A future high-energy $e^+e^-$ collider will allow studies of quarkonium
production in two-photon collisions. In a Higgs factory mode, it would
afford new opportunities to make precision measurements of the Higgs
couplings to Standard Model particles. The Higgs decay to
$J/\psi+\gamma$ provides, perhaps, the only realistic means with which
to probe the $Hc\bar c$ coupling. Higgs decays to $J/\psi+Z$ and
$\Upsilon+Z$ are other quarkonium-related avenues for probing Higgs
couplings.

Double-charmonium production has been the focus of interesting
studies at the $B$ factories and it will be investigated further  at
Belle~II. By taking advantage of the high luminosity at Belle~II,
one should be able to observe final states, such as $J/\psi+J/\psi$,
that were not produced at sufficient rates to have been observed
previously at $B$ factories. Accurate measurements of the cross
sections for exclusive charmonium production will provide motivation for
developing understanding and control of large logarithms of
$p_T^2/m_q^2$ that appear in the theoretical predictions. 

\section{Quarkonium Spectroscopy$^{\,}$\protect\footnotemark}
\footnotetext{Authors: Eric Braaten, Estia 
Eichten, Stephen Lars Olsen, Todd K.~Pedlar}

\subsection{Introduction}

The strongly interacting particles in the Standard Model 
are quarks and gluons. The strongly interacting particles in Nature 
are mesons and baryons.  In the Standard Model, 
quarks and gluons are related to mesons and baryons by the 
long-distance regime of QCD, which remains the least understood aspect 
of the theory.  Since first-principle calculations 
using lattice QCD are still not practical for many
long-distance phenomena, 
a number of models motivated by the color structure of QCD 
have been proposed.  However, so far at least, 
predictions of these QCD-motivated models that pertain to 
the spectrum of hadrons have a less-than-stellar record.  
In spite of decades of experimental searches, 
unambiguous examples of hadrons predicted by these models, 
such as pentaquarks, the $H$-dibaryon, 
mesons with exotic $J^{PC}$ quantum numbers, etc., have not been found.  
On the other hand, a number of states that do not fit into the 
conventional picture of hadrons {\it have} been found.  
A compelling, unified understanding of these new states 
has not yet emerged.  This gap between quark/gluon theory 
and meson/baryon observations remains a major deficiency 
in our current level of understanding of elementary particle processes.

For the reasons discussed above, theoretical and experimental 
investigations of the spectroscopy of quarkonium and 
quarkonium-like hadrons have had a profound influence on 
the development of our current level of understanding of 
the relation between quarks/gluons and hadrons.  
It is essential (and probably inevitable) that these investigations 
will continue to be pursued, with the goals of ultimately 
establishing a reliable QCD-motivated theoretical model that 
can accurately describe measured phenomena and experimental 
observations of the hadron spectra predicted by this model.  
Insights gained from these activities will likely have applications 
far beyond the limited area of quarkonium physics.

One of the most basic properties of QCD is its spectrum:
the list of particles that are stable 
or at least sufficiently long-lived to be observed as resonances.
The elementary constituents in QCD are quarks ($q$), 
antiquarks ($\bar q$), and gluons ($g$), and QCD requires that
they be confined into color-singlet clusters called hadrons.
The most stable hadrons are the clusters predicted by the quark model:
mesons ($q \bar q$), baryons ($q q q$), and antibaryons.
Hundreds of meson resonances and dozens of baryons resonances  
have been observed, most of which are conventional mesons
and conventional baryons.  Until recently, no other types of clusters 
had been unambiguously identified.
The general expectation in the high energy physics community has been 
that other types of clusters were probably broad and were probably
strongly mixed with conventional mesons and baryons,
in which case identifying them would be a complicated problem in hadron physics.
This expectation has been shattered by recent discoveries of 
charged quarkonia in the 
bottomonium ($b \bar b$) and charmonium ($c \bar c$) sectors of QCD.
The discoveries of these manifestly exotic tetraquark mesons have
exposed an embarrassing gap in our qualitative understanding of QCD.

The discoveries of the charmonium and bottomonium tetraquarks
are the culmination of a decade of surprising experimental results 
on heavy quarkonium.
It began with the discovery of the $X(3872)$  
by the Belle Collaboration in September 2003.
This $c \bar c$ meson has  decays that severely violate isospin symmetry,
disfavoring its identification as conventional charmonium.
It continued with the discovery of the $Y(4260)$
by the $\babar$ Collaboration in July 2005.
This $c \bar c$ meson has $J^{PC}$ quantum numbers $1^{--}$, 
but it is produced very weakly in $e^+ e^-$ annihilation.  
Over the past decade, many additional $c \bar c$ mesons have been observed.
Those with masses below the open-charm threshold fit 
neatly into the predicted charmonium multiplets. 
However most of those above the open-charm 
threshold have no compelling conventional charmonium
assignments. The list of these neutral $c \bar c$ mesons, 
which have been labeled $X$, $Y$, or $Z$,
has grown to more than a dozen states.
They are candidates for exotic clusters,
such as charmonium hybrids ($c \bar c g$) 
or charmonium tetraquarks ($c \bar c q \bar q$).
However none of these neutral $c \bar c$ mesons is manifestly exotic,
with $J^{PC}$ or flavor quantum numbers that are incompatible with 
conventional charmonium.  Thus, despite their unexpected properties,
one could not exclude these states from simply being complicated 
manifestations of the strong interactions between hadrons.

The recent discoveries of charged quarkonium-like mesons in the  
$b \bar b$ and $c \bar c$ sectors eliminates the possibility of dismissing the 
$XYZ$ mesons as complicated manifestations of the strong interactions.
In November 2011, the Belle Collaboration discovered 
the $Z_b^+(10610)$
and  $Z_b^+(10650)$, whose decays into $\Upsilon \, \pi^+$ 
reveal their constituents to be $b \bar b u \bar d$.
In April 2013, the BESIII Collaboration discovered 
the $Z_c^+(3900)$, 
whose decays into $J/\psi\, \pi^+$ reveal its constituents
to be $c \bar c u \bar d$.  
This state was almost immediately confirmed by the Belle Collaboration.
Despite being above the thresholds for decays into pairs of heavy-light mesons,
these charged tetraquark mesons are relatively narrow.
They are a glimpse into a sector of the QCD spectrum that has remained hidden 
for all these decades.

The observation of many neutral $XYZ$ mesons with unexplained properties
and the discoveries of the charged $Z_b$ and $Z_c$ mesons 
have revealed a serious gap in our qualitative understanding of QCD.
The challenge of understanding the $Q \bar Q$ mesons above 
the open-heavy-flavor threshold has now become urgent.  
Calculations from first principles 
using lattice QCD should eventually be capable of revealing 
the spectrum of these mesons.  However understanding their properties 
will probably also require the development of phenomenological descriptions
that are well motivated by QCD.

Fortunately experiments at the intensity frontier and at the energy frontier
of high energy physics are well positioned to address this problem. 
Electron-positron colliders in Beijing
(BEPC II), which is now running in the charmonium region,
and in Japan (Super KEK-B), 
which will begin running in the bottomonium region in 2016,
provide clean environments for the study of heavy quarkonia 
and quarkonium-related states.  At the Large Hadron Collider,
the ATLAS, CMS, and LHCb detectors have already contributed to
quarkonium spectroscopy using proton-proton collisions.
They should be able to do much more after the beam energy upgrade 
in 2015 and after a subsequent luminosity upgrade at a later date.
The dedicated $B$-physics experiment LHCb has a particularly large
potential for discovery in the realm of quarkonium spectroscopy.
Finally, the PANDA experiment at the FAIR facility in Darmstadt, 
will utilize antiproton-proton annihilation 
in the charm-anticharm threshold region
to study charmonium and other neutral $c \bar c$ mesons 
beginning around 2019.
With the wealth of additional clues provided by all these experiments,
the emergence of a theoretical understanding of the quarkonium spectrum 
above the open-heavy-flavor threshold is inevitable.

\subsection{Experimental Results} 

\begin{sidewaystable}
\caption{
New $c\bar{c}$ and $b\bar{b}$ mesons above the open-heavy-flavor threshold. 
The {\bf bold} states have been observed by
at least two independent experiments
and with significance greater than 5$\sigma$.
The masses $M$ and widths $\Gamma$ are
weighted averages from the Experiments,
with uncertainties added in quadrature.
In the $J^{PC}$ column,
a question mark (?) indicates an unmeasured value.
For charged states, $C$ is that of a neutral isospin partner.
In the Process column, the decay modes are indicated in parentheses
and ellipses (...) indicate an inclusive measurement.
For each Experiment, the statistical significance is given in
standard deviations (\#$\sigma$) unless it is not provided (np). 
These Tables are adapted and updated from a table
in Ref.~\cite{Eidelman:2012vu} that was prepared in May 2012.
 } 
\setlength{\tabcolsep}{0.21pc}
\label{tab:Q-QbarI}
\begin{center}
\begin{tabular}{lccclll}
\hline\hline
\rule[10pt]{-1mm}{0mm}
 State & $M$~(MeV) & $\Gamma$~(MeV) & $J^{PC}$ & Process~(decay mode) & 
     Experiment~(\#$\sigma$) & 1$^{\rm st}$~observation \\[0.7mm]
\hline
\rule[10pt]{-1mm}{0mm}
$X(3823)$& 3823.1$\pm$1.9 & $<24$ &
    $?^{?-}$
    & $B \to K + (\chi_{c1}\gamma)$ &
    Belle~\cite{Bhardwaj:2013rmw}~(3.8) & Belle~2013 \\[0.7mm]
$\bm{X(3872)}$& 3871.68$\pm$0.17 & $<1.2$ &
    $1^{++}$
    & $B \to K + (J/\psi\, \pi^+\pi^-)$ &
    Belle~\cite{Choi:2003ue,Choi:2011fc}~(12.8), 
    \babar~\cite{Aubert:2008gu}~(8.6) & Belle~2003 \\[0.7mm]
& & & & $p\bar p \to (J/\psi\, \pi^+\pi^-)+ ...$ &
    CDF~\cite{Acosta:2003zx,Abulencia:2006ma,Aaltonen:2009vj}~(np), 
    \DZero~\cite{Abazov:2004kp}~(5.2) & \\[0.7mm]
& & &   & $B \to K + (J/\psi\, \pi^+\pi^-\pi^0)$ &
    Belle~\cite{Abe:2005ix}\footnote{\label{notinc-footnote} Not included in the
    averages for $M$ and $\Gamma$.}~(4.3),
    \babar~\cite{delAmoSanchez:2010jr}\textsuperscript{\ref{notinc-footnote}}~(4.0) & \\[0.7mm]
& & & & $B \to K + (D^0 \bar D^0 \pi^0)$ &
    Belle~\cite{Gokhroo:2006bt,Aushev:2008su}\textsuperscript{\ref{notinc-footnote}}~(6.4), 
    \babar~\cite{Aubert:2007rva}\textsuperscript{\ref{notinc-footnote}}~(4.9) & \\[0.7mm]
& & & & $B \to K + (J/\psi\, \gamma)$ &
    Belle~\cite{Bhardwaj:2011dj}\textsuperscript{\ref{notinc-footnote}}~(4.0), 
    \babar~\cite{Aubert:2006aj,Aubert:2008rn}\textsuperscript{\ref{notinc-footnote}}~(3.6)& \\[0.7mm]
& & & & $B \to K + (\psi(2S)\, \gamma)$ &
    \babar~\cite{Aubert:2008rn}\textsuperscript{\ref{notinc-footnote}}~(3.5),
    Belle~\cite{Bhardwaj:2011dj}\textsuperscript{\ref{notinc-footnote}}~(0.4) & \\[0.7mm]
& & & &  $pp \to (J/\psi\, \pi^+\pi^-)+ ...$   & 
    LHCb~\cite{Aaij:2012lhcb}~(np) & \\[1.9mm]
$\bm{X(3915)}$ & $3917.5\pm1.9$ & 20$\pm 5$ & $0^{++}$ &
    $B \to K + (J/\psi\, \omega)$ &
    Belle~\cite{Abe:2004zs}~(8.1),
    \babar~\cite{Aubert:2007vj}~(19) & Belle~2004 \\ [0.7mm]
     & & & & $e^+e^- \to e^+e^- + (J/\psi\, \omega)$ &
    Belle~\cite{Uehara:2009tx}~(7.7),
    \babar~\cite{delAmoSanchez:2010jr,    Lees:2012xs}(7.6~)& \\[1.9mm]
$\bm{\chi_{c2}(2P)}$ & $3927.2\pm2.6$ & 24$\pm$6 & $2^{++}$ &
     $e^+e^-\to e^+e^- + (D\bar{D})$ &
     Belle~\cite{Uehara:2005qd}~(5.3), 
     \babar~\cite{:2010hka}~(5.8) & Belle~2005 \\ [1.9mm]
$X(3940)$ & $3942^{+9}_{-8}$ & $37^{+27}_{-17}$ & $?^{?+}$ &
     $e^+e^- \to J/\psi + (D^* \bar D)$ &
     Belle~\cite{Abe:2007sya}~(6.0) & Belle~2007 \\ [0.7mm]
&&&& $e^+e^- \to J/\psi + (...)$ &
     Belle~\cite{Abe:2007jn}~(5.0) \\ [1.9mm]
$\bm{G(3900)}$ & $3943\pm21$ & 52$\pm$11 & $1^{--}$ &
     $e^+e^- \to \gamma + (D \bar D)$ &
     \babar~\cite{Aubert:2006mi}~(np),
     Belle~\cite{Pakhlova:2008zza}~(np)
     & \babar~2007 \\ [1.9mm] 
$Y(4008)$ & $4008^{+121}_{-\ 49}$ & 226$\pm$97 & $1^{--}$ &
     $e^+e^- \to \gamma + (J/\psi\, \pi^+\pi^-)$ &
     Belle~\cite{Belle:2007sj}~(7.4)~
     & Belle~2007 \\[1.9mm]
$\bm{Y(4140)}$ & $4144.5\pm2.6$  & $15^{+11}_{-\ 7}$ & $?^{?+}$ &
     $B \to K + (J/\psi\, \phi)$ &
     CDF~\cite{Aaltonen:2009tz,Aaltonen:2011at}~(5.0),
     CMS~\cite{Yetkin:2013iza}~($>$5) & CDF~2009 \\[1.9mm]
$X(4160)$ & $4156^{+29}_{-25} $ & $139^{+113}_{-65}$ & $?^{?+}$ &
     $e^+e^- \to J/\psi + (D^* \bar D^*)$ &
     Belle~\cite{Abe:2007sya}~(5.5) & Belle~2007 \\[1.9mm]
     \hline
\end{tabular}
\end{center}
\end{sidewaystable}

\begin{sidewaystable}
\caption{
New $c\bar{c}$ and $b\bar{b}$ mesons above the open-heavy-flavor threshold
(continuation of Table~\ref{tab:Q-QbarI}). } 
\setlength{\tabcolsep}{0.21pc}
\label{tab:Q-QbarII}
\begin{center}
\begin{tabular}{lccclll}
\hline\hline
\rule[10pt]{-1mm}{0mm}
 State & $M$~(MeV) & $\Gamma$~(MeV) & $J^{PC}$ & Process~(decay mode) & 
     Experiment~(\#$\sigma$) & 1$^{\rm st}$ observation \\[0.7mm]
\hline
\rule[10pt]{-1mm}{0mm}
$\bm{Y(4260)}$ & $4263^{+8}_{-9}$ & 95$\pm$14 & $1^{--}$ &
     $e^+e^- \to \gamma + (J/\psi\, \pi^+\pi^-)$ &
     \babar~\cite{Aubert:2005rm,Aubert:2008ic}~(8.0),
     CLEO~\cite{He:2006kg}~(5.4)
     & \babar~2005 \\ [0.7mm]
           & & & & &  Belle~\cite{Belle:2007sj}~(15) &\\[0.7mm]
& & & & $e^+e^-\to (J/\psi\, \pi^+\pi^-)$ & CLEO~\cite{Coan:2006rv}~(11)& \\[0.7mm]
& & & & $e^+e^-\to (J/\psi\, \pi^0\pi^0)$ & CLEO~\cite{Coan:2006rv}~(5.1) & \\[1.9mm]
$Y(4274)$ & $4274.4^{+8.4}_{-6.7}$ & $32^{+22}_{-15}$ & $?^{?+}$ &
     $B\to K + (J/\psi\, \phi )$ &
     CDF~\cite{Aaltonen:2011at}~(3.1) & CDF~2010 \\[1.9mm]
$X(4350)$ & $4350.6^{+4.6}_{-5.1}$ & $13.3^{+18.4}_{-10.0}$ & 0/2$^{++}$ &
     $e^+e^-\to e^+e^- \,(J/\psi\, \phi)$ &
     Belle~\cite{Shen:2009vs}~(3.2) & Belle~2009 \\ [1.9mm]
$\bm{Y(4360)}$ & $4361\pm13$ & 74$\pm$18 & $1^{--}$ &
     $e^+e^-\to\gamma + (\psi(2S)\, \pi^+\pi^-)$ &
     \babar~\cite{Aubert:2006ge}~(np),
     Belle~\cite{:2007ea}~(8.0) & \babar~2007 \\ [1.9mm]
$X(4630)$ & $4634^{+\ 9}_{-11}$ & $92^{+41}_{-32}$ & $1^{--}$ &
     $e^+e^-\to\gamma\, (\Lambda_c^+ \Lambda_c^-)$ &
     Belle~\cite{Pakhlova:2008vn}~(8.2)  & Belle~2007 \\ [1.9mm]
$Y(4660)$ & 4664$\pm$12 & 48$\pm$15 & $1^{--}$ &
     $e^+e^-\to\gamma + (\psi(2S)\, \pi^+\pi^-)$ &
     Belle~\cite{:2007ea}~(5.8)  & Belle~2007 \\ [0.7mm]
\hline
$\bm{Z_c^+(3900)}$ & $3898\pm 5$ & $51\pm 19$ & $1^{?-}$ &
     $Y(4260) \to \pi^- + (J/\psi\, \pi^+)$ &
     BESIII~\cite{Ablikim:2013mio} (np), Belle~\cite{Liu:2013dau} (5.2)  
     &  BESIII~2013\\[1.9mm]
 &  & & &  $e^+e^- \to \pi^- + (J/\psi\, \pi^+)$ &
     Xiao {\em et al.}~\cite{Xiao:2013iha}\footnote{Not included in the averages
     for $M$ and $\Gamma$.}  (6.1)&  \\[1.9mm]
$Z_1^+(4050)$ & $4051^{+24}_{-43}$ & $82^{+51}_{-55}$ & ?&
     $ B \to K + (\chi_{c1}(1P)\, \pi^+)$ &
     Belle~\cite{Mizuk:2008me}~(5.0),
     \babar~\cite{Lees:2011ik}~(1.1) & Belle~2008 \\[1.9mm]
$Z_2^+(4250)$ & $4248^{+185}_{-\ 45}$ &
     177$^{+321}_{-\ 72}$ &?&
     $ B \to K + (\chi_{c1}(1P)\, \pi^+)$ &
     Belle~\cite{Mizuk:2008me}~(5.0),
     \babar~\cite{Lees:2011ik}~(2.0) & Belle~2008 \\[1.9mm]
$Z^+(4430)$ & $4443^{+24}_{-18}$ & $107^{+113}_{-\ 71}$ & ?&
     $B \to K + (\psi(2S)\, \pi^+)$ &
     Belle~\cite{Choi:2007wga,Mizuk:2009da}~(6.4),
     \babar~\cite{:2008nk}~(2.4) & Belle~2007 \\[1.9mm]
\hline\hline
$Y_b(10888)$ & 10888.4$\pm$3.0 & 30.7$^{+8.9}_{-7.7}$ & $1^{--}$ &
      $e^+e^- \to (\Upsilon(nS)\, \pi^+\pi^-)$ &
      Belle~\cite{Chen:2008pu,Abe:2007tk}~(2.0)& Belle~2010 \\[0.7mm]
\hline
$Z_{b}^+(10610)$ & 10607.2$\pm$2.0 & 18.4$\pm$2.4 & $1^{+-}$ &
       $\Upsilon(5S) \to \pi^- + (\Upsilon(nS)\,\pi^+)$,~$n=1,2,3$~ &
      Belle~\cite{Adachi:2011XXX,Bondar:2011pd}~(16) & Belle~2011 \\[0.7mm]
 &  &  &  &
      $\Upsilon(5S) \to \pi^- + (h_b(nP)\,\pi^+)$,~$n=1,2$ & 
       Belle~\cite{Adachi:2011XXX,Bondar:2011pd}~(16) &  \\[1.9mm]
$Z_{b}^+(10650)$ & 10652.2$\pm$1.5 & 11.5$\pm$2.2 & $1^{+-}$ &
       $\Upsilon(5S)\to\pi^- + (\Upsilon(nS)\,\pi^+)$,~$n=1,2,3$ &
    Belle~\cite{Adachi:2011XXX,Bondar:2011pd}~(16)& Belle~2011 \\[0.7mm]
 &  &  &  &
       $\Upsilon(5S) \to \pi^- + (h_b(nP)\,\pi^+)$,~$n=1,2$ & 
       Belle~\cite{Adachi:2011XXX,Bondar:2011pd}~(16)& \\[1.9mm]
\hline\hline
\end{tabular}
\end{center}
\end{sidewaystable}

Ten years have passed since the announcement by the Belle Collaboration 
at the 2003 Lepton-Photon Conference of the discovery of a new 
charmonium-like state lying above the open-charm threshold, 
referred to as $X(3872)$~\cite{Choi:2003ue,Choi:2011fc}.  
The $X(3872)$ was only the first of many quarkonium-like states 
above the open-heavy-flavor threshold that were
discovered over the course of the last decade.
These states are listed in Tables~\ref{tab:Q-QbarI} and \ref{tab:Q-QbarII}. 
The list is separated into four sections. 
The first section is the list of new neutral $c \bar c$ mesons, which
overpopulate the region 
from the open-charm threshold up to approximately 4700 MeV 
as compared to the number of expected conventional charmonium states.
The second section is the list of charged $c \bar c$ mesons, 
which are manifestly exotic tetraquarks.
The third and fourth sections are
a single new neutral $b \bar b$ meson
and the two charged $b \bar b$ mesons.
The charged $Q \bar Q$ mesons presumably belong to isospin triplets
that include a neutral meson.  The neutral mesons $Z_c^0(3900)$ and 
$Z_b^0(10610)$ have been observed, but they are not listed in the Tables.
Additional information about many these states can be found in 
several reviews \cite{Godfrey:2008nc,Barnes:2009zza,Pakhlova:2010zz}.

We proceed to summarize the new quarkonia below the open-heavy-flavor 
threshold that have been discovered in the last decade.
We then discuss selected aspects of the new quarkonium-like states 
above the open-heavy-flavor thresholds.

\subsubsection{Quarkonia below the open-heavy-flavor thresholds}

\noindent

Prior to 2002, the knowledge of the spectra of narrow, below-threshold 
charmonium and bottomonium states was fairly limited.  
In the case of bottomonium, the masses, full widths, 
and dilepton partial widths 
of the triplet-$S$ states were known.
The masses of the triplet-$P$ states were also known.  
Dipion decays among triplet-$S$ states had been observed, 
and the radiative cascades involving
the triplet-$P$ states had also been measured, 
although with relatively poor precision.  
Additionally, a number of
radiative decays of the ground-state triplet-$S$, 
the $\Upsilon(1S)$, had been measured. 
Glaringly absent from the bottomonium spectrum, however, 
were the singlet-$S$ and singlet-$P$ states, 
as were the triplet-$D$ states,
whose ground states were expected to lie 
between the ground and first excited triplet-$P$ levels.  

Charmonium was on somewhat  firmer footing.  All the masses 
and full widths of the triplet-$S$ and triplet-$P$ states were known. 
The radiative transition rates among them were known to good precision, 
and many hadronic decays 
of each of these states had been observed.  In addition, 
the ground-state singlet-$S$, the $\eta_c(1S)$, had also
been observed by several experiments, though its full width 
was a hotly disputed matter.  As in the case 
of bottomonium, the singlet-$P$ state had not yet been observed, 
though hints had been obtained by two 
experiments.  

In the intervening decade, the picture for both 
the charmonium spectrum and the bottomonium spectrum
below their respective open-heavy-flavor thresholds
has become much clearer.  
In charmonium, work by E835, CLEO, BES/BESIII, Belle and $\babar$ have
firmly established all three below-threshold singlet 
states, $\eta_c(1S)$, $\eta_c(2S)$ and $h_c(1P)$.  
Their masses and widths are now
well measured, and many of their key decay modes are known.  
Each of these states has subsequently served as a vehicle 
for discovery of other new states, particularly 
in bottom meson decays and in studies of resonances lying above 
the open-charm thresholds.  

In bottomonium, the $\Upsilon(1D)$  was first observed by CLEO in 
2002~\cite{Bonvicini:2004yj}, and finally confirmed 
by $\babar$ in 2010~\cite{delAmoSanchez:2010kz}.  
This was the first firmly established $D$-wave quarkonium state. 
In 2008, the $\eta_b(1S)$ was observed by $\babar$~\cite{Aubert:2008ba} 
and CLEO~\cite{Bonvicini:2009hs}. 
The announcement of weak evidence for the $h_b(1P)$ 
at $\babar$~\cite{Lees:2011zp} was followed up the discovery of strong signals 
for both $h_b(1P)$ and $h_b(2P)$ in dipion transitions 
from $\Upsilon(5S)$ at Belle~\cite{Adachi:2011ji}.  Subsequent analyses of
the principal radiative decays of $h_b(1P)$ and $h_b(2P)$ 
yielded a more precise measurement of the
mass of $\eta_b(1S)$ and very strong evidence 
for its radial excitation, $\eta_b(2S)$~\cite{Mizuk:2012pb}.  

\subsubsection{$X(3872)$}
\label{sec:X3872}

The $X(3872)$ was first observed~\cite{Choi:2003ue} 
by the Belle Collaboration decaying to $J/\psi\, \pi^+\pi^-$ and
produced opposite a charged kaon in the decay of charged $B$ mesons.
The state was quickly confirmed by inclusive
production in $p \bar p$ collisions by CDF~\cite{Acosta:2003zx} 
and \DZero~\cite{Abazov:2004kp} and also by  
$\babar$ in the discovery channel~\cite{Aubert:2008gu}. 
The observation by Belle~\cite{Bhardwaj:2011dj} 
and $\babar$~\cite{Aubert:2006aj} of the
decay mode $J/\psi\, \gamma$ established the $C$-parity as $+$. 
A 2006 analysis by the CDF collaboration of the $J/\psi\, \pi^+\pi^-$
decay mode~\cite{Abulencia:2006ma} reduced the possible $J^{PC}$ 
quantum numbers to $1^{++}$ and $2^{-+}$.  
A 2010 analysis by $\babar$~\cite{delAmoSanchez:2010jr} 
of the $J/\psi\, \pi^+\pi^-\pi^0$
decay mode indicated a moderate preference for $2^{-+}$ over $1^{++}$.  
In May 2013, LHCb published the results of a full five-dimensional angular 
analysis of the $J/\psi\, \pi^+\pi^-$ decays of $X(3872)$~\cite{Aaij:2012lhcb},
which ruled out $2^{-+}$ by over $8\sigma$ compared to $1^{++}$.  
The $1^{++}$ quantum numbers are those of the conventional 
$^3P_1$ charmonium state $\chi_{c1}(2P)$,
but the $X(3872)$ has other properties that 
make such an assignment problematic.

A very important property of the $X(3872)$ is that its
mass is extremely close to the $D^{*0} \bar D^0$ threshold.
Its energy relative to that threshold is measured to be 
$- 0.3 \pm 0.4$~MeV.  The $1^{++}$ quantum numbers of $X(3872)$ 
imply that it has an $S$-wave coupling to $D^* \bar D$.
As discussed in Sec.~\ref{sec:molecule},
the universal properties of $S$-wave near-threshold resonances
then guarantee that the $X(3872)$ must be a charm-meson molecule 
whose constituents are a superposition of $D^{*0} \bar D^0$ 
and $D^0 \bar D^{*0}$.
One simple universal prediction for such a loosely bound $S$-wave molecule
is that the root-mean-square separation of its constituents 
scales with its binding energy $E_X$ as $E_X^{-1/2}$ \cite{Braaten:2003he}.
If the binding energy of the $X(3872)$ is 0.3~MeV,
the root-mean-square separation of its constituent charm mesons 
is predicted to be a remarkable 5~fm.

The property of the $X(3872)$ that was initially most surprising
is that its decays revealed a dramatic violation of isospin symmetry.
In the discovery mode $J/\psi\, \pi^+\pi^-$, the $\pi^+\pi^-$ system
is compatible with the decay of a virtual $\rho$, 
so the decay mode must have isospin 1. 
In the decay mode $J/\psi\, \pi^+\pi^- \pi^0$, the $\pi^+\pi^- \pi^0$ system
is compatible with the decay of a virtual $\omega$, 
so the decay mode must have isospin 0. 
The decay modes $J/\psi\, \pi^+\pi^-$ and $J/\psi\, \pi^+\pi^- \pi^0$
were observed to have comparable branching fractions,
indicating a severe violation of isospin symmetry.
This can be explained by the fact that the mass 
of the $X(3872)$ is extremely close to the $D^{*0} \bar D^0$ threshold
but about 8~MeV below the $D^{*+} D^-$ threshold,
provided the isospin of $X(3872)$ would be 0 if it were not 
so close to these thresholds.
The branching fractions into 
$J/\psi\, \pi^+\pi^-$ and $J/\psi\, \pi^+\pi^- \pi^0$ are comparable
in spite of the much stronger coupling of $X(3872)$ to $J/\psi\, \omega$,
because the decay rate of $\rho$ into $\pi^+\pi^-$ 
is much larger than that of $\omega$ into $\pi^+\pi^- \pi^0$
and because the decay into $J/\psi\, \omega$ proceeds 
through the tail of the $\omega$ resonance.

\subsubsection{New $1^{--}$ mesons}
\label{sec:one--}

Several new $J^{PC} = 1^{--}$ mesons have been observed 
by CLEO, Belle and \babar, spanning a range of masses 
between 3900 MeV and 4700 MeV.  
All of them were produced via Initial State Radiation (ISR), 
which fixes their quantum numbers to be that of the photon. 
Most of these states were observed in dipion transitions 
to either $J/\psi$ or $\psi(2S)$ -- 
the only exceptions being $G(3900)$~\cite{Aubert:2006mi},
which was observed decaying to $D\bar{D}$, and $X(4630)$~\cite{Pakhlova:2008vn}, 
which was observed decaying to 
$\Lambda_c^+ \Lambda_c^-$.

The most well-established of these states, the $Y(4260)$, 
was observed in ISR decaying to 
$J/\psi\, \pi^+\pi^-$ by \babar~\cite{Aubert:2005rm,Aubert:2008ic},
CLEO~\cite{He:2006kg} and Belle~\cite{Belle:2007sj}.  
It was later observed through direct $e^+e^-$ annihilation 
on resonance by CLEO~\cite{Coan:2006rv} -- an analysis which also
produced evidence of the $Y(4260)$ decaying by a  
neutral dipion transition to $J/\psi\, \pi^0\pi^0$.
Thanks to the large Belle data sample at $\Upsilon(4S)$ 
and the dedicated BES III running at $\sqrt{s} = 4260$ MeV, 
the statistics accumulated for the $Y(4260)$ are
sufficient for detailed studies of resonant 
substructure in its decays to $J/\psi\, \pi^+\pi^-$.
As noted below in Sec.~\ref{sec:charged}, 
these studies have revealed additional unexpected states,
in particular the charmonium tetraquark $Z_c^+$.

The $Y(4260)$ resonance is a candidate for a 
charmonium hybrid.  One of the basic characteristics of a 
quarkonium hybrid is that its $Q \bar Q$ wave function 
is strongly suppressed near the origin. 
The production rate in $e^+ e^-$ annihilations of a hybrid 
with quantum numbers $1^{--}$ is therefore strongly
suppressed.  The $Y(4260)$ is produced so weakly in $e^+ e^-$ annihilation 
that it appears as a small local maximum in the cross section 
near the deep minimum between the conventional charmonium states 
$\psi(4160)$ and $\psi(4415)$.

In bottomonium, a candidate for an additional $J^{PC}=1^{--}$ resonance
above the open-bottom threshold was observed by Belle at a mass just above
that of the $\Upsilon(5S)$.  In 2008, Belle published the analysis of 
$\Upsilon(5S)\rightarrow \Upsilon(nS)\, \pi^+\pi^-$~\cite{Abe:2007tk}, 
which indicated branching fractions that were
two orders of magnitude larger
than expected.  In 2010, Belle published the results of a study 
of the cross sections for
$e^+e^-\rightarrow \Upsilon(nS)\, \pi^+\pi^-$ 
as a function of $\sqrt{s}$~\cite{Chen:2008pu}, in which it
was found that the peak cross section for this process 
lies at 10.89~GeV, rather 
than at 10.86~GeV, where the total hadronic cross section peaks.  
This result has been interpreted as arising from a 
distinct $1^{--}$ resonance 
very close to $\Upsilon(5S)$~\cite{Hambrock:2013tpa}, 
but this has not yet been confirmed.  
Since it is not predicted by quark potential models, 
this $Y_b(10890)$ resonance, if confirmed, is likely
to have an unconventional quark structure,
such as a bottomonium hybrid, tetraquark or molecule.

\subsubsection{New positive $C$-parity mesons}

Positive $C$-parity charmonium-like mesons can be produced 
at $e^+e^-$ colliders via the two-photon fusion process
$\gamma \gamma \rightarrow c\bar{c}$ or the double
charmonium process $e^+e^- \rightarrow J/\psi + c\bar{c}$.
They can also be produced by more complicated processes, 
such as the decays of $B$ mesons or inclusive production 
in hadron collisions, in which case their $C$-parity
can be revealed by their decay products,
such as $J/\psi$ plus a light vector meson $\omega$ or $\phi$.

The $X(3915)$ has the most substantial pedigree, having been observed 
by both Belle~\cite{Abe:2004zs,Uehara:2009tx} 
and $\babar$~\cite{Aubert:2007vj,delAmoSanchez:2010jr,Lees:2012xs} 
and having been produced both in $B$ decays 
and in $\gamma\gamma$ fusion.  In all these cases, 
the $X$ is observed in its decay to $J/\psi\, \omega$.    
Spin-parity measurements by the $\babar$ Collaboration
prefer $0^{++}$.  This makes the $X(3915)$
a candidate for the $\chi_{c0}(2P)$, but this assignment 
is disfavored by its near mass degeneracy 
with the candidate $\chi_{c2}(2P)$ state (see paragraph below) 
and its narrow width, 
which suggests that its decays into $D \bar D$ must be suppressed.

Both Belle~\cite{Uehara:2005qd} and $\babar$~\cite{:2010hka} 
have observed a state at a mass of 3927~MeV
produced by $\gamma\gamma$ fusion, which decays to $D\bar{D}$.  
The analysis of spin and $C$-parity favors an assignment of $2^{++}$.
Since there is no strong evidence disfavoring this assignment, 
we have denoted this state $\chi_{c2} (2P)$.   

The $X(3940)$ has been observed only by the Belle Collaboration
in $e^+ e^-$ collisions, but using two different methods.
It has been observed through the exclusive final state 
$J/\psi\, D^* \bar D$ as a resonance in $D^* \bar D$~\cite{Abe:2007sya}.  
It has also been observed through the inclusive final state $J/\psi + X$
as a peak in the recoil momentum distribution for $J/\psi$~\cite{Abe:2007jn}.

The CDF Collaboration has demonstrated that general-purpose 
detectors at hadron colliders can also contribute to 
quarkonium spectroscopy by their discovery of the $Y(4140)$
through its decay into $J/\psi\, \phi$~\cite{Aaltonen:2009tz,Aaltonen:2011at}.
The $Y(4140)$ has been confirmed at the LHC 
by the CMS Collaboration~\cite{Yetkin:2013iza}.

\subsubsection{Charged $Q \bar Q$ mesons}
\label{sec:charged}

The observations of charged resonances above the open-heavy-flavor 
thresholds that couple strongly to charmonium or bottomonium
have led to significant interest in both the experimental 
and theoretical communities.  Such resonances, decaying to
a heavy quarkonium state and a charged light-quark meson  
are manifestly exotic, requiring a minimal quark content of 
$Q \bar Q q \bar q$.
Given the existence of this new category of tetraquark hadrons,
the planned data sets at the upgraded Super KEK-B collider
as well as the current and future data sets from the LHC experiments
offer significant discovery potential.

The Belle Collaboration has reported the observation  
of three charged charmonium-related resonances with 
statistical significances greater than $5\sigma$:  
the $Z_1^+(4050)$ and $Z_2^+(4250)$ were observed 
in the $\chi_{c1}(1P)\,  \pi^+$ final state~\cite{Mizuk:2008me} 
and $Z^+(4430)$ was observed in 
$\psi(2S)\, \pi^+$~\cite{Choi:2007wga,Mizuk:2009da}.  
$\babar$ searched for similar structures in the
same final states~\cite{Lees:2011ik,:2008nk}, 
but was unable to observe statistically 
significant excesses for any of them.  

More recently, in decays of the
$Y(4260)$, both the BES III~\cite{Ablikim:2013mio}  
and the Belle~\cite{Liu:2013dau} Collaborations announced 
the observation of a new charged charmonium-like resonance, 
$Z_c^+(3900)$, which decays to $J/\psi\, \pi^+$.  
The Belle Collaboration utilized a sample of $Y(4260)$ mesons
produced using Initial State Radiation (ISR), 
while the BES Collaboration's result was based on 
a sample of $e^+e^-$ annihilations recorded on the peak of 
the $Y(4260)$.  Both experiments observed signals with
greater than $5\sigma$ significance.  
In addition, an analysis based on CLEO-c data taken at 
$\sqrt{s} = 4170 $ MeV claimed a $6.0\sigma$ 
observation of the
 $Z_c(3900)$ in the 
$J/\psi\, \pi^+$ final state~\cite{Xiao:2013iha}, 
although the mass measurement is incompatible by nearly $3\sigma$ with 
the Belle and BES results.  In addition, the same study 
presented $3\sigma$ evidence for a neutral isospin partner $Z_c^0$
in the $J/\psi\, \pi^0$ decay channel at a similar mass. 

Charged bottomonium-like mesons were discovered by the 
Belle Collaboration in 2011. 
In an effort to understand the unexpectedly large rate of production 
of the singlet-$P$ states $h_b(2P)$ and $h_b(1P)$ observed
in $\pi^+\pi^-$ transitions from $\Upsilon(5S)$, 
the Belle Collaboration studied the resonant substructure in these 
transitions~\cite{Adachi:2011XXX,Bondar:2011pd}.  
Two significant peaks, denoted $Z_b(10610)$ 
and $Z_b(10650)$, were observed in the invariant mass of 
$\Upsilon(nS)\,  \pi^+$ and in that of 
$h_b(nP)\, \pi^+$ as well.  In fact, it was observed 
that the rate of production of $h_b(nP)$ in $\pi^+\pi^-$ transitions
from $\Upsilon(5S)$ was fully accounted for by charged pion cascades 
involving the $Z_b$ states, and large fractions of the 
transitions to the lower $\Upsilon(nS)$ states occur through the $Z_b$ as well.  
Both charged $Z_b$ states were observed in dipion
transitions to $\Upsilon(1S,2S,3S)$ and $h_b(1P,2P)$ 
with consistent mass and width measurements for all of the five final 
states.  Belle subsequently obtained 4.9$\sigma$ evidence 
for the neutral isospin partner of the $Z_b(10610)$ 
in $\pi^0\pi^0$ transitions from $\Upsilon(5S)$ 
to $\Upsilon(2S)$~\cite{Adachi:2012im}.

The $Z_b(10610)$ and $Z_b(10650)$ states lie very close to the $BB^*$ 
and $B^*B^*$ thresholds, respectively, and therefore
raise the speculation that the two states are, in fact, 
molecular states of these bottom meson pairs.  
This speculation is supported
by the subsequent observation by Belle~\cite{Adachi:2012cx}
of large rates of decay of the $Z_b$ states into the expected $B^{(*)}$ meson 
pairs.  
 
The $Z_c^+(3900)$ is the only charged quarkonium resonance 
that has been observed in more than one experiment. 
Since the only known production
mechanism of the $Z_b$ states is $\Upsilon(5S)$ decays
and as the Belle Collaboration 
has the only substantial $\Upsilon(5S)$ data sample, 
the observations of $Z_b(10610)$ and $Z_b(10650)$
are likely to remain unconfirmed until
further data are taken on the $\Upsilon(5S)$ resonance,
unless these states can be observed inclusively in 
$\Upsilon\, \pi^+$ decays in the LHC experiments.
The confirmation of these exotic $b \bar b$ mesons is clearly important.
The discovery of additional tetraquark mesons
may provide essential clues for our understanding 
of this new aspect of the QCD spectrum.
We therefore anticipate exciting results 
when new experiments come online in the next
decade, providing data samples sufficiently large 
to explore this new territory with
much increased precision.  

\subsection{Phenomenological Models}

The properties of the neutral and charged $XYZ$ states
are presumably described by QCD.  
The ultimate challenge to theory
is to derive those properties directly from QCD.
A more modest challenge for theory is to develop a phenomenological 
description of these states motivated by QCD that 
accurately reproduces the properties of the states that have
already been observed and
predicts which additional states should be observed 
and which ones should not.
Thus far, theory has failed to even meet this modest challenge.

Below, we describe briefly the phenomenological models 
of the $XYZ$ mesons that have been proposed.
Such a model can be 
specified most simply by identifying its colored constituents 
and how they are clustered within the meson.
A more detailed description might specify the forces 
between the constituents.
Thus far none of the phenomenological models that have been proposed 
has produced a compelling explanation for the 
neutral and charged $XYZ$ states that have been observed.

\subsubsection{Conventional quarkonium}

The only constituents in a conventional quarkonium 
are the heavy quark and antiquark ($Q \bar Q$).
In a quark potential model, the $Q$ and $\bar Q$ interact through
a potential $V(r)$ that can be approximated by a color-Coulomb potential 
proportional to $1/r$ at short distances 
and to a confining potential that is linear in $r$ at long distances.
The quarkonium spectrum in the quark potential model
consists of spin-symmetry multiplets:
$S$-wave multiplets with $J^{PC}$ quantum numbers
$\{ 0^{-+}, 1^{--} \}$,
$P$-wave multiplets $\{ 1^{+-}, (0,1,2)^{++} \}$,
D-wave multiplets $\{ 2^{-+}, (1,2,3)^{--} \}$, 
F-wave multiplets $\{ 3^{-+}, (2,3,4)^{--} \}$, etc.
The splittings between multiplets are approximately the same  
in charmonium and bottomonium.  
The splittings within spin-symmetry multiplets are
approximately 3 times smaller in bottomonium than in charmonium.

Quark potential models provide a very accurate description of 
the quarkonium states below their open-heavy-flavor threshold.
In the absence of large mixings with other types of mesons,
they should also provide a good description of conventional
quarkonium states above the open-heavy-flavor threshold.
In addition to the quarkonium masses, potential models provide
predictions for the radiative transition rates
between quarkonium states.
Models for their hadronic transition rates based on the 
multipole expansion of QCD have also been developed.
Predictions for charmonium states above the open-charm threshold 
are given in Refs.~\cite{Eichten:2004uh,Barnes:2005pb,Eichten:2007qx}.

\subsubsection{Meson molecules}
\label{sec:molecule}

If the mass of a quarkonium-like meson is sufficiently close 
to the open-heavy-flavor threshold 
for a pair of heavy-light mesons, it is at least plausible 
to interpret the pair of mesons as constituents of the 
quarkonium-like meson.
For such an interpretation to be plausible, the widths of the 
constituents must be narrower than that of the meson.
The charm mesons that are narrow enough to be plausible 
constituents are the $S$-wave mesons $D$, $D^*$, $D_s$, and $D_s^*$
and the $P$-wave mesons $D_1$, $D_2^*$, $D_{s0}^*$, $D_{s1}$, and $D_{s2}^*$.
Pairs of these mesons with zero net strangeness 
define 25 thresholds between 3730~MeV and 5146~MeV. 
Thus the mass of any given $c \bar c$ meson in this region
has a high probability of randomly
being within 50~MeV of one of the thresholds.

The interactions between the heavy-light mesons at sufficiently 
low energies is due to pion exchange, 
where the tensor term
in the pion-exchange potential is singular at short distance.
In channels where the pion-exchange potential is attractive,
whether the mesons are bound into molecules or unbound 
depends crucially on the short-distance region.  
For reasonable choices of a 
short-distance cutoff, there are bound states near threshold 
for $D^* \bar D$ with isospin 0 and $J^{PC} = 0^{-+}$ and $1^{++}$
and for $D^* \bar D^*$ with isospin 0 and $J^{PC} = 0^{++}$,
$0^{-+}$, $1^{+-}$, and $2^{++}$ \cite{Tornqvist:1993ng}.
The corresponding bottom meson pairs are predicted to form
molecules in these channels that are more strongly bound. 
The pattern of quantum numbers for the molecules can be changed 
by taking into account the potentials from the exchange 
of other mesons.

If the mass of a quarkonium-like meson is extremely close to the 
threshold for a pair of mesons and if it has an $S$-wave coupling 
to them, it can be identified rigorously as a loosely bound 
molecule consisting of those mesons \cite{Braaten:2003he}.
The binding energy must be much less than the energy scale 
set by the range of the pion-exchange interactions 
between the pair of mesons, which is about 10~MeV 
from pion exchange for charm meson pairs 
and about 3~MeV for bottom meson pairs.
In this case, the molecule has universal properties 
that are determined by its binding energy and its width.
The quintessential example of such a molecule is the $X(3872)$, 
as discussed in Sec.~\ref{sec:X3872}.

\subsubsection{Quarkonium hybrids}
\label{sec:hybrid}

A quarkonium hybrid has constituents $Q \bar Q g$,
where $g$ is a constituent gluon.  
Since the color charge of a gluon is octet, 
the $Q \bar Q$ pair is also in a color-octet state.
In the lowest-energy multiplets of quarkonium hybrids, 
the constituent gluon $g$ is a magnetic gluon 
that can be assigned the quantum numbers $J^{PC} = 1^{+-}$
and can be interpreted as a vector particle in a $P$-wave orbital.
If the $Q \bar Q$ pair is in an $S$-wave state,
the states of the hybrid form the spin-symmetry multiplet 
$\{ 1^{--}, (0,1,2)^{-+} \}$.
If the $Q \bar Q$ pair is in a $P$-wave state,
the states of the hybrid form a supermultiplet 
that can be decomposed into the spin-symmetry multiplets
$\{ 1^{++}, (0,1,2)^{+-} \}$,
$\{ 0^{++}, 1^{+-} \}$,  and
$\{ 2^{++}, (1,2,3)^{+-} \}$.
The quantum numbers $1^{-+}$, $0^{+-}$, and $2^{+-}$ are exotic:
they are not possible for a conventional quarkonium 
that consists of $Q \bar Q$ only.
As described in Sec.~\ref{sec:lattice},
calculations of the $c \bar c$ meson spectrum using lattice QCD 
have found that the lowest charmonium hybrids fill out 
an $S$-wave multiplet and a $P$-wave supermultiplet.
As described in Sec.~\ref{sec:latticeNRQCD},
calculations of the $b \bar b$ meson spectrum using lattice 
NRQCD without dynamical quarks have found that the lowest 
bottomonium hybrid spin-symmetry multiplets are
$\{ 1^{--}, (0,1,2)^{-+} \}$, $\{ 1^{++}, (0,1,2)^{+-} \}$,
and $\{ 0^{++}, 1^{+-} \}$,
followed by an excited $\{ 1^{--}, (0,1,2)^{-+} \}$ multiplet.

An alternative model of the quarkonium hybrids
is provided by the Born-Oppenheimer approximation,
in which the $Q$ and $\bar Q$ move adiabatically in
the presence of gluon and light-quark fields 
whose energy levels are those generated by static 
$Q$ and $\bar Q$ sources.
In the absence of dynamical light quarks, the Born-Oppenheimer potentials 
can be calculated straightforwardly using lattice QCD.
As described in Sec.~\ref{sec:BO}, if the Born-Oppenheimer
is applied to bottomonium without light quarks,
the lowest spin-symmetry multiplets are
$\{ 1^{--}, (0,1,2)^{-+} \}$ and $\{ 1^{++}, (0,1,2)^{+-} \}$, 
followed by radial excitations of these two multiplets, and
then by $\{ 0^{++}, 1^{+-} \}$.

The lowest Born-Oppenheimer potential for a quarkonium hybrid
has a minimum for a $Q \bar Q$ separation 
of about 0.3~fm, and it is repulsive at short distances.
As a consequence, the wave function at the origin
for the $Q \bar Q$ in the quarkonium hybrid is very small.
A $1^{--}$ quarkonium hybrid should therefore be
produced very weakly in $e^+ e^-$ annihilation 
compared to a conventional $1^{--}$ quarkonium.

One model-independent prediction for decays of a quarkonium hybrid 
is that decay into a pair of $S$-wave mesons 
is suppressed \cite{Kou:2005gt,Close:2005iz}.
The dominant decays of charmonium hybrids were therefore
previously expected to be into an $S$-wave and $P$-wave charm-meson pair, 
provided these states are kinematically accessible.
The discoveries of tetraquark $Q \bar Q$ mesons provide 
previously unanticipated decay channels for quarkonium hybrids.

\subsubsection{Compact tetraquarks}

A compact tetraquark has constituents $Q \bar Q q \bar q$ 
that are in overlapping orbitals.
In the simplest quark potential models for tetraquarks, the four quarks
interact through pair-wise potentials between the six quark pairs.
In such models, the mass of a tetraquark must be below the 
thresholds for pairs of heavy-light mesons ($Q \bar q$ and $\bar Q q$)
with appropriate quantum numbers and below the 
thresholds for a quarkonium ($Q \bar Q$) 
plus a light  ($\bar q q$) meson
with appropriate quantum numbers \cite{Vijande:2007ix}.  
If the tetraquark is above either of these thresholds, 
there is no potential barrier 
that prevents it from falling apart into a pair of mesons.
In more complicated quark potential models,
there can also be 3-quark and 4-quark potentials 
that may be able to stabilize tetraquarks 
above the meson pair thresholds \cite{Vijande:2011im}.

\subsubsection{Diquarkonium tetraquarks}

One possibility for substructure in a $Q \bar Q q \bar q$ 
tetraquark is that its constituents are clustered into diquarks 
$Q q$ and $\bar Q \bar q$ \cite{Drenska:2010kg}.  
Their color states are antitriplet and triplet, respectively.  
The spin quantum number of a diquark can be 0 or 1.
In the naive diquark model,
the only degrees of freedom come from the 
spins and flavors of the quarks inside the diquarks
and from the relative motion of the diquarks, 
whose orbitals are determined by 
the orbital angular momentum and radial quantum numbers.
The tetraquark energy levels are then completely determined
by group theory in terms of the orbital energies 
and the spin-spin interaction strengths, which
are phenomenological parameters.

The lowest energy states of the tetraquark are those where 
the diquark pair is in an $S$-wave state. 
If we consider only the light quark flavors $u$ or $d$,
the $Q \bar Q$ tetraquark spectrum consists of 
degenerate isosinglet and isovector
multiplets with $J^P$ quantum numbers $0^+$, $0^+$, $1^+$, $1^+$, 
$1^-$, and $2^+$ \cite{Drenska:2010kg}.  
The $J^{PC}$ quantum numbers of the
neutral tetraquarks are $0^{++}$, $1^{++}$, $1^{+-}$, 
and $2^{++}$, with 4, 4, 2, and 2 states, respectively.
If we include the $s$ quark, there are additional 
isodoublet and isosinglet states.
If we also consider orbital angular momentum excitations 
and radial excitations of the diquark pair,
there is a vast proliferation of predicted states.

\subsubsection{Hadro-quarkonium}
Hadro-quarkonium consists of a color-singlet $q \bar q$ pair 
bound to a compact core that is a color-singlet 
$Q \bar Q$ pair \cite{Dubynskiy:2008mq}.
The binding can be attributed to a color analog of the van der Waals force
between neutral atoms.  
An essentially equivalent model is a molecule consisting of  
a light meson bound to a quarkonium.
Hadro-quarkonium is motivated primarily by the fact that 
most of the $XYZ$ states have been observed through their hadronic
transitions to a single charmonium state
by emitting a specific light vector meson.
This is explained naturally if the charmonium state 
and the light meson are already present as 
constituents of the $XYZ$ meson.

\subsubsection{Born-Oppenheimer tetraquarks}

The structure of a $Q \bar Q q \bar q$ tetraquark could be
similar to that of a quarkonium hybrid 
in the Born-Oppenheimer approximation, 
in which the $Q$ and $\bar Q$ move adiabatically in
the presence of gluon and light-quark fields 
whose energy levels are those generated by static 
$Q$ and $\bar Q$ sources \cite{Braaten:2013boa}.
In the case of tetraquarks, 
the gluon and light-quark fields have non-singlet flavor quantum numbers.
A simple predictive assumption for the tetraquark 
Born-Oppenheimer potentials is that they differ only 
by an offset in energy
from the corresponding quarkonium hybrid potentials.
This implies that the pattern of spin-symmetry multiplets 
for the tetraquarks is similar to that for the quarkonium hybrids.

\subsection{Theoretical approaches within QCD}

QCD presumably provides a fundamental description of the 
$Q \bar Q$ mesons above the open-heavy-flavor threshold.
In this section, we describe theoretical approaches to the problem 
that are directly based on QCD.
 
\subsubsection{Lattice QCD}
\label{sec:lattice}

The only well-developed systematically improvable method for 
calculating observables of QCD that are inherently nonperturbative 
is lattice gauge theory.
This method involves discretizing the quark and gluon fields 
on a Euclidean space-time lattice, evaluating the functional integrals
over the Grassmann quark fields analytically to get a determinant, 
and then evaluating the functional integrals over the gluon fields 
using Monte Carlo methods.  This method can only be applied directly
to observables that can be extracted from Euclidean correlation functions 
of the quark and gluon fields.  For such observables, 
all the errors in the calculation can be quantified.
There are statistical errors from the limited sample of gluon field 
configurations and also from the determination of the QCD parameters, 
which are the coupling constant $\alpha_s$ and the quark masses.
There are also systematic errors 
from the extrapolation of the lattice volume to infinity, 
the extrapolation of the lattice spacing to zero, 
and the extrapolation of the light quark masses to their physical values.

The mass of the charm quark is small enough that lattice gauge theory,
with presently available computational resources,
can be applied directly to $c \bar c$ mesons.
The most extensive studies of the $c \bar c$ meson spectrum 
above the $D \bar D$ threshold have been carried out by 
Dudek, Edwards, Mathur and Richards \cite{Dudek:2007wv}
and extended by the 
Hadron Spectrum Collaboration \cite{Liu:2012ze}.
The most recent published calculations
used an anisotropic lattice with $24^3 \times 128$ sites
and a spatial lattice spacing of about 0.12~fm.
Their gauge field configurations were generated using dynamical 
$u$, $d$, and $s$ quarks, with the $s$ quark having its physical 
mass and the $u$ and $d$ quarks unphysically heavy, 
corresponding to a pion mass of about 400~MeV.
On a cubic lattice, there are 20 channels analogous to the 
$J^{PC}$ quantum numbers in the continuum.
For each of the 20 lattice $J^{PC}$ channels,
the Hadron Spectrum Collaboration calculated the $c \bar c$ meson spectrum
from the Euclidean time dependence of the cross-correlators
for a set of operators whose number ranged from 4 to 26,
depending on the channel.  The operators included ``hybrid'' 
operators constructed out of the $c$ quark field 
and the gluon field strength and ``charmonium'' operators 
operators constructed out of the $c$ quark field and 
other combinations of covariant derivatives.
They considered two lattice volumes
and verified that their results were insensitive to the lattice volume.
Since they only considered one value for the lattice spacing
and one value for the $u$ and $d$ masses,
they could not quantify the systematic errors associated 
with the lattice spacing or the light quark masses.

The Hadron Spectrum Collaboration identified 46 states 
in the $c \bar c$ meson spectrum 
with high statistical precision \cite{Liu:2012ze}. 
These states had spins $J$  as high as 4 and masses as high as 4.6~GeV.  
Four of the states had exotic quantum numbers that were not possible 
for a pure $c \bar c$ state: $1^{-+}$, $0^{+-}$, and $2^{+-}$.
In some of the nonexotic $J^{PC}$ channels, there were several 
states that could not be accommodated by the multiplets for
conventional charmonium.  In particular, a total of 6 states 
were identified in both the $1^{--}$ and $1^{+-}$ channels.
The states that are more strongly excited by charmonium operators are 
plausible candidates for conventional charmonium.
They fill out complete $1S$, $2S$, $3S$, $1P$, $2P$, $1D$, and $1F$ multiplets
and there are also candidates for the $3P$ and $1G$ multiplets.
The states that are more strongly excited by hybrid operators are 
plausible candidates for charmonium hybrids.
They can be assigned to the heavy-quark spin-symmetry multiplets
$\{ 1^{--}, (0,1,2)^{-+} \}$,
$\{ 1^{++}, (0,1,2)^{+-} \}$,
$\{ 0^{++}, 1^{+-} \}$, and
$\{ 2^{++}, (1,2,3)^{+-} \}$ \cite{Liu:2012ze}.
Since the probabilities for light-quark constituents in 
a conventional quarkonium or in a quarkonium hybrid are expected to be small,
it is plausible that their results give the correct pattern 
for the spectrum of charmonium and charmonium hybrids in QCD.

Definitive calculations of the spectrum of $c \bar c$ mesons 
with all systematic errors quantified would require much more 
extensive calculations.  The calculations of Ref.~\cite{Liu:2012ze}
would have to be repeated at several smaller lattice spacings 
in order to extrapolate to zero lattice spacing.
They would have to be repeated for several smaller pion masses 
in order to extrapolate to the physical $u$ and $d$ masses.
The list of operators would have to be expanded to include 
``tetraquark'' ($c \bar c q \bar q$) operators that are constructed 
out of light-quark fields as well as the $c$ quark field.
A tetraquark operator that factors into 
color-singlet $c \bar q$ and $\bar c q$ operators
has an enhanced amplitude for exciting 
scattering states consisting of a pair of charm mesons,
which have a discrete spectrum on the lattice.
One complication that could become increasingly severe 
as the $u$ and $d$ quark masses are decreased is 
the effect of 3-hadron scattering states consisting of 
a $c\bar c$ meson and a pair of light mesons.

The masses of the one or two lightest $c \bar c$ mesons 
in each of 13 $J^{PC}$ channels have also been calculated 
by Bali, Collins, and Ehmann \cite{Bali:2011rd}.
They used isotropic lattices with 
$16^3 \times 32$ sites and $24^3 \times 48$ sites
and lattice spacings of 0.115~fm and 0.077~fm, respectively.
Their gauge field configurations were generated using dynamical 
$u$ and $d$ quarks with equal masses that correspond to a pion mass 
of 1010~MeV, 400~MeV, or 280~MeV.
In each $J^{PC}$ channel, the masses of the two lightest 
$c \bar c$ mesons were extracted from the cross-correlators 
of 3 operators.  For most of the $J^{PC}$ channels, the operators 
were charmonium operators, but the operators for the $2^{+-}$ channel 
were hybrid operators.  Comparison with the results of 
Ref.~\cite{Liu:2012ze} suggest that several of the states should
be interpreted as charmonium hybrids, namely the states with the 
exotic quantum numbers $1^{-+}$ and $2^{+-}$ 
and the first excited states for $2^{-+}$ and $3^{+-}$.
In Ref.~\cite{Bali:2011rd}, the mixing between $c \bar c$ mesons 
and pairs of charm mesons were also studied in the channels 
$0^{-+}$, $1^{--}$, and $1^{++}$.
 
The radiative transition rates between $Q \bar Q$ mesons 
can also be calculated using lattice QCD.
They can be extracted from the Euclidean correlation functions 
of three operators, one of which is the electromagnetic current operator.
The first such calculations for excited $c \bar c$ mesons 
were carried out by Dudek, Edwards, and Thomas 
using lattice QCD without dynamical quarks \cite{Dudek:2009kk}.
They used an anisotropic lattice with $12^3 \times 48$ sites
and a lattice spacing of about 0.1~fm.
They calculated the electric dipole transition rate
between each of the four lowest-energy $1^{--}$ states and the $\chi_{c0}$.
For the fourth $1^{--}$ state, the transition rate is very small, 
consistent with its identification as a charmonium hybrid.
They also calculated the magnetic dipole transition rate
between each of the four lowest-energy $1^{--}$ states and the $\eta_c$.

\subsubsection{Lattice NRQCD}
\label{sec:latticeNRQCD}

Since the mass of the $b$ quark mass is larger than that of the 
$c$ quark by about a factor of 3, lattice QCD calculations 
for $b \bar b$ mesons require a lattice spacing that is 3 times smaller 
to obtain the same accuracy as for $c \bar c$ mesons.
The resulting increase in the number of lattice points by a 
factor of $3^4$ puts lattice QCD calculations for $b \bar b$ mesons 
beyond the reach of the computational power that is currently available.
An alternative is to use lattice NRQCD.  Nonrelativistic QCD 
is an effective field theory for QCD in which the heavy quark 
is treated nonrelativistically. Lattice NRQCD has been very successful
in calculating the properties of the bottomonium states 
below the $B \bar B$ threshold. 
(For a recent example, see Ref. \cite{Dowdall:2011wh}.)
Since the $b$ and $\bar b$ remain nonrelativistic in the 
$b \bar b$ mesons above the $B \bar B$ threshold, 
NRQCD should also be applicable to those mesons.

Lattice NRQCD without any dynamical quarks
has been used by Juge, Kuti, and Morningstar to calculate 
the energies of the lightest bottomonium hybrid mesons \cite{Juge:1999ie}.
They used an anisotropic lattice with $15^3 \times 45$ sites
and a spacial lattice spacing of about 0.11~fm.
They used the Euclidean time-dependence of the correlator 
of a single hybrid operator in each of four $J^{PC}$ channels
to determine the energies of the hybrid states.
The ground-state hybrid spin-symmetry multiplet was determined to be
$\{ 1^{--}, (0,1,2)^{-+} \}$.  The lowest-energy excited multiplets
are $\{ 1^{++}, (0,1,2)^{+-} \}$ and $\{ 0^{++}, 1^{+-} \}$,
followed by an excited $\{ 1^{--}, (0,1,2)^{-+} \}$ multiplet.

\subsubsection{Born-Oppenheimer approximation}
\label{sec:BO}

Because the mass of a heavy quark is large,
the $Q$ and $\bar Q$ in a $Q \bar Q$ meson move slowly in response 
to the gluon and light-quark fields, while, in comparison,
the responses of the gluon and light-quark fields to the motion 
of the $Q$ and $\bar Q$ are almost instantaneous.
These features are exploited in the Born-Oppenheimer approximation. 
The relative motion of the $Q$ and $\bar Q$ is described by 
the Schroedinger equation with Born-Oppenheimer (B-O) potentials $V_n(r)$
defined by the energy levels of the gluon and light-quark fields
in the presence of static $Q$ and $\bar Q$ sources
separated by a distance $r$. The energy levels 
for the gluon and light-quark fields can be 
specified by the quantum number $+\Lambda$ or $-\Lambda$, 
where $\Lambda = 0,1,2,\ldots$ (or $\Lambda = \Sigma,\Pi,\Delta,\ldots$), 
for the projection of their total angular momentum $\bm{J}_{\rm light}$
on the $Q \bar Q$ axis and by their $CP$ quantum number
$\eta = \pm 1$ (or $\eta = g,u$).
The sign of $\pm \Lambda$ is relevant only for $\Lambda = \Sigma$.
The B-O potentials can therefore be labeled
$\Sigma_g^\pm$, $\Sigma_u^\pm$, $\Pi_g$, $\Pi_u$, \ldots.
The centrifugal potential in the Schroedinger equation is proportional to  
$(\bm{L} - \bm{J}_{\rm light})^2$, where $\bm{L}$ is the sum of 
$\bm{J}_{\rm light}$ and the orbital angular momentum of the $Q \bar Q$
pair.  The raising and lowering operators $J_{\rm light}^+$ 
and $J_{\rm light}^-$ in the $\bm{L} \cdot\bm{J}_{\rm light}$ term 
introduce couplings between the B-O potentials.
In the leading Born-Oppenheimer approximation, 
these coupling terms are neglected
and $L$ is a good quantum number.  The distinct energy levels 
from solutions to the Schroedinger equation 
for a specific B-O
potential can therefore be labeled $nL$, where $n$ is a 
radial quantum number and $L = S, P, D, \ldots$.
A systematic expansion around the leading Born-Oppenheimer approximation 
that takes into account nonadiabatic effects 
from the motion of the $Q \bar Q$ pair
can be developed by treating the 
$L^- J_{\rm light}^+$ and $L^+ J_{\rm light}^-$ terms 
in the centrifugal potential as perturbations.

In the leading Born-Oppenheimer approximation, 
each of the energy levels $nL$ for the 
Schroedinger equation with a specific B-O potential
corresponds to a multiplet of $Q \bar Q$ mesons.
In the flavor-singlet sector, the energy levels in the
ground-state B-O potential $\Sigma_g^+$ can be interpreted as
conventional quarkonia while those in the
excited B-O potentials $\Pi_u$, $\Sigma_u^-$, \ldots can be interpreted as
quarkonium hybrids.  In the flavor-nonsinglet sectors,
the energy levels in the B-O potentials can be interpreted 
as tetraquark mesons.

The lowest flavor-singlet Born-Oppenheimer potentials 
have been calculated by Juge, Kuti, and Morningstar using lattice QCD
without any dynamical quarks \cite{Juge:1999ie}.
Their finest lattices were $10^3 \times 30$ and $14^3 \times 56$,
with spacial lattice spacings of about 0.19~fm and 0.23~fm, respectively.
The ground-state potential $\Sigma_g^+$ behaves like the 
phenomenological potential of quark-potential models. 
The most attractive of the excited B-O potentials is the $\Pi_u$ potential, 
which has a minimum near 0.3 ~fm. This potential is repulsive 
and linear in $r$ at long distances, and it is repulsive at short distances,
where it approaches the energy of gluon and light-quark fields
in the presence of a static color-octet $Q \bar Q$ source.
The solutions to the Schroedinger equation for the $b \bar b$ pair
in the excited B-O potentials reveal that the lowest energy level for 
bottomonium hybrids is the $P$-wave level $\Pi_u(1P)$, 
whose supermultiplet consists of two degenerate spin-symmetry multiplets: 
$\{ 1^{--}, (0,1,2)^{-+} \}$ and $\{ 1^{++}, (0,1,2)^{+-} \}$. 
The next lowest energy levels are the radially excited $P$-wave level 
$\Pi_u(2P)$ and the $S$-wave level $\Sigma_u^-(1S)$,
which consists of a single spin-symmetry multiplet $\{ 0^{++}, 1^{+-} \}$.
If the nonadiabatic couplings between B-O potentials were taken into account,
the $\Pi_u(nP)$ supermultiplets would be split 
and the energy ordering of the $\Pi_u(2P)$ and $\Sigma_u^-(1S)$ levels 
would presumably be reversed, so that they would agree with the results 
of the lattice NRQCD calculations of bottomonium hybrids 
described in Sec.~\ref{sec:latticeNRQCD}.

In the presence of static $Q$ and $\bar Q$ sources,
the lowest-energy state of the gluon and light-quark fields 
need not be localized near the sources.
The localized fields can be accompanied by additional light hadrons.
These additional hadrons complicate the calculation of a B-O
potential, which should really be defined as the minimal energy for
{\it localized} gluon and light-quark fields with the appropriate quantum
numbers.  This problem is not so severe in the absence of light quarks,
because the only light hadrons are glueballs, 
which have rather large masses.  The problem is more serious 
if there are dynamical light quarks, and it becomes increasingly 
severe as the masses of the $u$ and $d$ quarks are decreased 
to their physical values.  It may be possible to overcome this problem, 
or at least ameliorate it, by using the cross-correlators 
of multiple operators to determine the B-O potentials.

\subsection{Scientific Opportunities} 

In this section, we describe the opportunities in the near 
and mid-term future for adding to our understanding
of the spectroscopy of quarkonia and related states.  
We describe, in roughly chronological order in terms of readiness for
data taking, the prospects for results in this arena from BESIII, 
Belle II, the LHC experiments, and PANDA.  

\subsubsection{BESIII @ BEPC} 
BESIII is the only experiment currently taking data 
using $e^+ e^-$ collisions.
It is expected to continue running for eight to ten years -- 
although its current specific run plan carries through only 2015. 
It has already made significant contributions 
to the study of charmonium and related states in the region 
above the open-charm threshold, 
both confirming neutral states 
and discovering new charged and neutral states.
Their decision to run
at the $Y(4260)$ has supplied interesting information 
well beyond what may have been expected, 
including the discovery of the charmonium tetraquark $Z_c^+$.

Among the possibilities that have good discovery potential 
would be continued running on the $Y(4260)$ resonance
and additional 
running on other higher $1^{--}$ resonances.  These 
priorities compete with the desires to run at the $\psi(3770)$ 
for studies of $D$ mesons and at the $\psi(4170)$ for studies
of $D_s$ mesons, but they should be given consideration.
Given the tantalizing possibility of the $Y(4260)$'s 
identification as a charmonium hybrid, more data taken
on resonance will certainly help support 
or deny this possibility. 
With a proven detector and an excellent accelerator facility that is 
currently operating, BES III alone has the immediate 
opportunity to add greatly to our understanding of the
$c \bar c$ meson spectrum above the open-charm threshold. 
We expect many new results from BES III in the near term.  

\subsubsection{Belle II @ Super KEK-B}

The KEK-B accelerator at KEK was dismantled after the completion of the 
operation of the Belle experiment in 2010, 
and currently is undergoing a substantial upgrade of the entire facility
to what will be known as Super KEK-B.  
The design luminosity of Super KEK-B is 
$8 \times 10^{35}~{\rm cm}^{-2}\, {\rm s}^{-1}$, 
nearly 40 times the peak 
instantaneous luminosity achieved by KEK-B.  
The plans are to begin running at the $\Upsilon(4S)$ resonance 
in 2016, and ultimately to collect an integrated luminosity
of 50 ${\rm ab}^{-1}$.  The Belle II detector also 
represents a substantial upgrade to the Belle detector system -- 
including new particle-ID 
systems, central drift chamber, and silicon and pixel detectors 
for increased vertexing capabilities.  

The Belle Collaboration has had remarkable success
in quarkonium spectroscopy, 
particularly in the study 
of states accessible through $B$ decays,
double charmonium production, and by means of 
Initial State Radiation (ISR), as well as those accessible in 
$e^+e^-$ annihilation near the $\Upsilon(5S)$.
This gives reason for optimism concerning the prospects 
for further success with the upgraded detector system 
and the much increased luminosities.

Many interesting states have been observed by Belle 
in running at the $\Upsilon(4S)$, but most, if not all of them, 
require significantly increased statistics in order to better 
characterize their properties and decays.  
With projected increases of factors of forty to fifty 
in integrated luminosity, the prospects for these studies 
(and new discoveries) are bright indeed.  In 
particular, we would also like to note the importance 
of running at the $\Upsilon(5S)$, and possibly above the resonance, 
for elucidating the character of the charged and neutral $Z_b$ 
tetraquark states, and transitions to lower bottomonia.   
The clean event environment provided by $e^+e^-$ annihilations 
offers great advantages for the study of both the tetraquark 
states and conventional bottomonia -- as well as discovery potential 
for new states higher in mass.  It would
be a great loss if a large increase in statistics at the $\Upsilon(5S)$ 
and above were not accumulated in Belle II.  

\subsubsection{LHC upgrade}
Currently the LHC is in a two-year shutdown for an upgrade to the RF 
and superconducting magnets, enabling not only an
increase in the center-of-mass energy 
but also a large increase in the instantaneous luminosity.  
Already, ATLAS, CMS and LHCb have made contributions 
to the study of some of the states we have discussed in this paper.
LHCb made the definitive determination of the quantum numbers $1^{++}$
of the $X(3872)$.  LHCb and CMS have made extensive measurements of the 
production rate of the $X(3872)$. 
CMS has confirmed the CDF observation of the $Y(4140)$ in the 
$J/\psi\, \phi$ decay mode.
Another indication of the capabilities of the LHC experiments is
provided by the conventional charmonium states $\chi_{bJ}(3P)$,
which were discovered by ATLAS and confirmed by \DZero\ and by LHCb.
With the planned increases in both energy and luminosity over the next decade, 
the LHC experiments promise to be even more
fruitful sources of new results in this area.

\subsubsection{PANDA @ FAIR}

PANDA, an experiment planned for the FAIR facility in Germany, 
offers a possibility not exploited since the experiments
E760 and E835 at Fermilab: antiproton-proton annihilation 
on resonance into charmonium states.   States of all $J^{PC}$
quantum numbers may be directly accessed by this technique, 
so long as they have large enough branching fractions to $p\bar{p}$.
To produce conventional charmonium in $p \bar p$ collisions,
all three of the quarks in the proton must annihilate
with antiquarks in the $\bar p$. 
Tetraquark mesons with constituents $c \bar c q \bar q$
can be produced in $p \bar p$ collisions by annihilating 
only two of the three quarks in the $p$.
Thus tetraquark charmonium states may couple more strongly to $p \bar p$
than conventional charmonium states with a similar energy.

PANDA is expected to begin data taking around 2019.
This gives other experiments, such as LHCb, 
the opportunity to discern which new states might 
couple strongly enough to $p\bar{p}$ to be ripe targets for
study at PANDA.

\section{Quarkonium Production$^{1,2}$}
\addtocounter{footnote}{1}\footnotetext{Authors: Geoffrey T.~Bodwin, Eric
Braaten, James Russ}
\addtocounter{footnote}{1}\footnotetext{More detailed accounts of technical 
aspects of quarkonium production and comparisons between theory and 
experiment can be found in 
Refs.~\cite{Brambilla:2004wf,Brambilla:2010cs,Bodwin:2012ft}}

\subsection{Introduction}

Quarkonium production rates have been studied intensely since the
discovery of the $J/\psi$. One aspect of quarkonium production that
makes it an interesting problem is the variety of different quarkonia
that can be studied. Nature has provided three sets of heavy-quarkonium
systems: charmonium ($c \bar c$), bottomonium ($b \bar b$), and $b \bar
c$ mesons. The most attractive targets for production measurements are
the narrowest states. Most of them are below the threshold for pairs of
heavy-light mesons, although there may also be exceptional, narrow
states above threshold, such as the $X(3872)$.

In the charmonium and bottomonium systems, the states whose production
rates are most easily measured at hadron colliders
are the spin-triplet S-wave states: the
$J/\psi$ and the $\psi(2S)$ in the charmonium system and the
$\Upsilon(1S)$, the $\Upsilon(2S)$, and the $\Upsilon(3S)$ in the
bottomonium system. They have significant decay modes into $\mu^+ \mu^-$
and $e^+ e^-$, which allow accurate measurements and provide useful
triggers. The next most easily measured states are the spin-triplet
P-wave states: the $\chi_{cJ}(1P)$ states ($J=0,1,2$) in the charmonium
system, and the $\chi_{bJ}(1P)$ and $\chi_{bJ}(2P)$ states in the
bottomonium system. They have radiative decays into a spin-singlet
S-wave state and a photon.  At $e^+ e^-$ colliders, the production 
rates of many of the other quarkonium states below the 
heavy-flavor threshold can be measured.

In the $b \bar c$ system, the only meson whose production rate has been
measured thus far is the spin-singlet ground state $B_c^-$, which has
been observed through its decays into $J/\psi\, \ell^- \bar\nu_\ell$
(Ref.~\cite{Abulencia:2006zu}) and its decays into $J/\psi\, \pi^\pm$
(Ref.~\cite{Aaltonen:2007gv}). Lattice QCD should provide a reliable
value for the partial width for this decay mode, which can be combined
with the lifetime measurement to determine a cross section. Measurements
of the relative branching ratios for the $B_c$ into hadronic modes would
give additional constraints on theoretical models of the decay process.
Measurements of the production rate of the $B_c$ are valuable in order
to test the predicted dependence of the rate on the heavy-quark mass
ratio.  Measurements of the production rates of any of the excited $b
\bar c$ mesons may be challenging.

An important consequence of the large mass of the heavy quark is the
suppression of nonperturbative (low-momentum-transfer) interactions that
change the spin state of the quarkonium. Consequently, quarkonium
polarizations in hard-scattering production processes are amenable to
perturbative theoretical analyses.  The polarizations of quarkonia are
therefore important observables with which to test quarkonium-production
theory.

\subsection{Importance of quarkonium production}

\subsubsection{Intrinsic importance}

The production rate of specific hadrons in high-energy collisions is a
fundamental problem in QCD that is important in its own right. The
hadrons whose production rate should be easiest to understand are heavy
quarkonia --- bound states consisting of a heavy quark and a heavy
antiquark. The large quark mass $m_Q$ implies that the creation of
the heavy quark and antiquark can be described using perturbative QCD.
The fact that heavy quarkonia are nonrelativistic bound states allows
the application of theoretical tools that simplify and constrain the
analyses of nonperturbative effects that are associated with their
formation. Hence, heavy quarkonium production provides a unique
laboratory in which to explore the interplay between perturbative and
nonperturbative effects in QCD.

Recent theoretical developments, which are described in
Sec.~\ref{sec:prodtheory}, have improved the prospects for a rigorous
and quantitative understanding of quarkonium production in the kinematic
region of large momentum transfer\footnote{\label{pT-footnote}
Throughout this paper, we use $p_T$ to denote the large momentum
transfer. If either of the colliding particles is a hadron, then $p_T$
denotes the quarkonium transverse momentum.  If both incoming particles
are leptons or photons, then $p_T$ denotes the quarkonium momentum in
the center-of-mass frame.} $p_T$. Experiments at the energy frontier can
measure quarkonium production at much larger $p_T$ than ever before,
providing definitive tests of this new theoretical framework.

\subsubsection{Laboratory for experimental analysis}

Some of the most important decay modes of heavy elementary
particles have analogs in quarkonium decays. For example, the decay $Z^0
\to \mu^+ \mu^-$ has the quarkonium analogs $\Upsilon(1S) \to \mu^+
\mu^-$ and $J/\psi \to \mu^+ \mu^-$. In a high-energy hadron collider,
it is possible to accumulate very large data samples of quarkonia. 
Analysis methods for decays of very heavy particles can therefore be
tested by making use of the analogous quarkonium decays.

An important example is the measurement of the polarization or spin
alignment of a particle.  The polarization is often measured by choosing
a polarization frame and measuring the polar-angle distribution with
respect to the polarization axis. However such a measurement, by itself,
does not provide the best control of systematic errors. The large data
samples of the $J/\psi$, the $\Upsilon(1S)$, and other quarkonium states
that were accumulated at the Tevatron were used to measure their
polarizations \cite{Acosta:2001gv,Abulencia:2007us,Abazov:2008aa}. These
measurements are also being pursued at the LHC
\cite{Chatrchyan:2012woa}. They are demonstrating that, in order to get
the best control over systematic errors, it is essential to measure the
polarization for various choices of the polarization frame and to use
frame-independent relations to constrain the measurements
\cite{Faccioli:2010kd}. The power of the frame-independent relations is
that they are sensitive to the presence of residual background
contributions, since backgrounds tend to transform differently between
reference frames than do decay products. These lessons have not yet been
exploited in measurements of the polarizations of heavy elementary
particles, such as the $W$, the $Z^0$, and the top quark. Furthermore,
these lessons may be useful in determining the spins of any new
heavy elementary particles that are discovered.

\subsubsection{Laboratory for QCD theory}
\label{sec:QCDlab}

Quarkonium production can be used as a laboratory to test theoretical
concepts that are central to the perturbative QCD theoretical program.
For example, perturbative QCD at high momentum transfer $p_T$ relies heavily
on resummations of large logarithms of $p_T$ to achieve accurate
predictions. Such resummations can be tested in quarkonium production at
large $p_T$, where the clean experimental signals for the
$J/\psi$ and the $\Upsilon(1S)$ that are provided by the
$\mu^+\mu^-$ decay channel allow high-statistics measurements in
which background contributions are under good control. Precise
measurements might allow tests of subleading contributions to
resummation formulas. 

Factorization theorems are the theoretical foundation for all
perturbative calculations in QCD. Quarkonium production could be used to
test the validity of factorization theorems at leading and
subleading powers of $p_T^2$. Such tests would improve our
understanding of the interplay of perturbative and nonperturbative
physics in QCD and might also lead to theoretical insights into the
applicability of factorization concepts to new processes.

Exclusive quarkonium production may afford a unique opportunity to
tackle the long-standing problem of endpoint logarithms in exclusive
processes in QCD. In exclusive quarkonium production, such logarithms
occur at scales of order $m_Q$ or higher, and, therefore, can be
analyzed entirely within perturbation theory. Analyses of endpoint
logarithms in quarkonium production might lead to insights into the
nature of nonperturbative endpoint logarithms in exclusive processes for
light hadrons and to methods that would allow one to organize and resum
them.

\subsubsection{Probes for new physics\label{sec:probes-new-physics}}

Quarkonium production can be used as a tool to probe for new physics. 
The production of the $J/\psi$ or the $\Upsilon(1S)$, 
followed by their 
decays to $\mu^+\mu^-$, provides a clean experimental signal 
that can be measured with high statistics.
If the theory of quarkonium production can be made sufficiently
precise, then discrepancies between theory and experiment
at large $p_T$ or at large $\sqrt{s}$ might signal physics 
beyond the Standard Model. 

Physics beyond the Standard Model could include bound states of new
heavy particles. New heavy-particle bound states are an essential
feature of most technicolor  models. There are supersymmetric extensions
of the Standard Model in which some of the SUSY partners are most easily
discovered through their bound states. Specifically, there are models in
which the stop or gluino can be most easily discovered through decays of
stoponium or gluino-onium into photons. The theoretical techniques
that are required for understanding quarkonium production in QCD are
relevant for understanding the production of nonrelativistic bound
states of new particles.

Decays of heavy elementary particles into quarkonia could be used to
study the couplings of those heavy particles to heavy quarks. For
example, one could measure quarkonium production in decays of the Higgs
particle. The decay of the Standard Model Higgs to $J/\psi+\gamma$ or
$\Upsilon+\gamma$ involves an interference between a direct process, in
which the Higgs decays to a $Q\bar Q$ virtual pair, and an indirect
process, in which a top or $W$ loop produces a pair of photons or
$Z^0$'s, one of which decays into a quarkonium \cite{Bodwin:2013gca}.
The direct processes, and, hence, the decay rates are sensitive to the
$HQ\bar Q$ coupling. Although it may be difficult to observe, the
decay of the Higgs into $J/\psi+\gamma$ is the only process, as far as
is known, that can be used to probe the $Hc\bar c$ coupling
directly at the LHC. One could also use the decay of the Higgs in the
channel $H\to Z+Z^*\to Z+J/\psi$ or $Z+\Upsilon$ to probe the $HZZ$
coupling \cite{Isidori:2013cla}.

\subsection{Theoretical Framework}
\label{sec:prodtheory}

The earliest attempts to describe  quarkonium production were the {\it
color-singlet model}
\cite{Kartvelishvili:1978id,Chang:1979nn,Berger:1980ni,Baier:1981uk} and
the {\it color-evaporation model} \cite{Fritzsch:1977ay,Halzen:1977rs},
whose origins go back almost to the discovery of charmonium. They were
superseded in the 1990's by the {\it nonrelativistic QCD (NRQCD)
factorization approach} \cite{Bodwin:1994jh}, which uses an effective
field theory to separate perturbative and nonperturbative effects of
QCD.  Although it has not been derived rigorously from QCD, NRQCD
factorization remains a theoretically and phenomenologically viable
description of quarkonium production. It is the default model for most
current experimental studies. A more recent theoretical development is
the {\it next-to-leading power (NLP) fragmentation approach}
\cite{Kang:2011zza,Kang:2011mg,Fleming:2012wy}.  It is believed to be
valid up to corrections that go as $m_Q^4/p_T^4$. Hence, its predictions
should become most accurate at high values of $p_T$ that are accessible
at the energy frontier.  These modern theoretical approaches to
quarkonium production are described in more detail below.

\subsubsection{NRQCD Factorization Approach\label{sec:nrqcd-fact}}

NRQCD is an effective field theory for the sector of QCD that includes a
heavy quark ($Q$) and a heavy antiquark ($\bar Q$) whose velocities in
the $Q\bar Q$ rest frame, denoted by $v$, are nonrelativistic ($v\ll
1$). The NRQCD factorization conjecture \cite{Bodwin:1994jh} states that
the inclusive cross section for producing a quarkonium state $H$ with
sufficiently large momentum transfer $p_T$ (see
footnote~\ref{pT-footnote}) can be written as a sum of products of
short-distance $Q\bar Q$ production cross sections times long-distance
{\it NRQCD matrix elements}:
\begin{equation}
d \sigma[A+B \to H + X] = \sum_n d \sigma[A+B \to (Q \bar Q)_n + X]~
\langle  {\cal O}_n^H \rangle.
\label{NRQCD-fact}
\end{equation}
The sum over $n$ extends over the color and angular-momentum states of
the $Q \bar Q$ pair. The short-distance cross sections $d \sigma$ are
essentially inclusive partonic cross sections for creating a $Q\bar
Q$ pair, convolved with parton distributions if the colliding particles
$A$ and $B$ are hadrons. They are process dependent and have perturbative
expansions in powers of $\alpha_s(m_Q)$. The NRQCD matrix element
$\langle{\cal O}_n^H \rangle$ is essentially the probability for a
$Q\bar Q$ pair in the state $n$ to evolve into the heavy quarkonium $H$.
These nonperturbative factors are process-independent constants that
scale with definite powers of the relative velocity $v$. Hence, the
NRQCD factorization formula is a double expansion in powers of
$\alpha_s$ and $v$. The inclusive annihilation decay rate of a
quarkonium state satisfies a factorization formula that is analogous
to Eq.~(\ref{NRQCD-fact}), but with different NRQCD matrix elements.

The NRQCD production matrix elements are vacuum expectation values of
four-fermion operators in NRQCD, but with a projection onto an
intermediate state of the quarkonium $H$ plus anything:
\begin{equation}
\langle {\cal O}_n^H \rangle=\langle 0|\chi^\dagger \kappa_n\psi 
\biggl(\sum_X |H+X\rangle\langle H+X|\biggr) \psi^\dagger 
\kappa'_n\chi|0\rangle.
\label{NRQCDme}
\end{equation}
Here, $\psi^\dagger$ and $\chi$ are two-component (Pauli) fields that
create a heavy quark and a heavy antiquark, respectively, and $\kappa_n$
and $\kappa_n'$ are direct products of Pauli and color
matrices.\footnote{ The color-octet NRQCD matrix elements also include
Wilson lines that run from the quark and antiquark fields to infinity
\cite{Nayak:2005rw}. For simplicity, we have omitted these Wilson lines
in Eq.~(\ref{NRQCDme}).} A key feature of NRQCD factorization is that
quarkonium production can occur through the creation of color-octet, as
well as color-singlet, $Q\bar Q$ pairs. The color-singlet
contributions at the leading order in $v$ correspond to the contributions
of the color-singlet model. Hence, the NRQCD factorization approach
contains all of the production processes of the color-singlet model, as
well as additional production processes that involve color-singlet
production at higher orders in $v$ and color-octet production. The
leading color-singlet NRQCD production matrix element in
Eq.~(\ref{NRQCD-fact}) is simply related to the leading color-singlet
NRQCD decay matrix element in the analogous factorization formula for
the annihilation decay rate of the quarkonium $H$. Hence, the leading
color-singlet NRQCD production matrix elements can be determined from
quarkonium electromagnetic decay rates (up to corrections of order
$v^4$). The color-octet NRQCD production matrix elements are treated as
phenomenological parameters.

The predictive power of NRQCD factorization comes from truncating the
expansion in $v$ so as to reduce the number of NRQCD matrix elements that
must be fixed by phenomenology.  The truncation in $v$ is more accurate
for bottomonium than for charmonium. ($v^2\approx 0.1$ for the
$\Upsilon(1S)$; $v^2\approx 0.23$ for the $J/\psi$.) The truncation in
$v$ can reduce the number of nonperturbative constants to just a few
for each quarkonium spin multiplet. The truncation for S-wave states
that is used in current phenomenology includes the NRQCD matrix elements
through relative order $v^4$. For a spin-triplet S-wave quarkonium state
$H$, such as the $J/\psi$ or the $\Upsilon(1S)$, the nonperturbative
factors are reduced to a single color-singlet matrix element
$\langle{\cal O}^{H}(^3S_1^{[1]})\rangle$, which is of leading order in
$v$, and three color-octet matrix elements,
$\langle{\cal O}^{H}(^1S_0^{[8]})\rangle$,
$\langle{\cal O}^{H}(^3S_1^{[8]})\rangle$, and
$\langle{\cal O}^{H}(^3P_J^{[8]})\rangle$,
which are of relative orders $v^3$, $v^4$, and $v^4$, respectively.
In the notations for these NRQCD operators, the quantities 
in parentheses are the angular-momentum ($^{2S+1}L_J$)
and color state (singlet or octet) of the $Q\bar Q$ pair that
evolves into the quarkonium state $H$.

\subsubsection{NLP Fragmentation Approach}

For very large momentum transfer\textsuperscript{\ref{pT-footnote}} $p_T$,
the inclusive cross section to produce a hadron can be simplified.
The contribution to the cross section at leading power in $p_T$
($1/p_T^4$ in $d \sigma/dp_T^2$) can be written as a sum of
single-parton production cross sections convolved with 
{\it single-parton fragmentation functions} \cite{Collins:1981uw}:
\begin{equation}
d \sigma[A+B \to H + X] = \sum_i d\hat\sigma[A+B\to i+X]\otimes D[i\to H] .
\label{1-ple-frag}
\end{equation}
The sum over $i$ extends over the types of partons (gluons, quarks, and
antiquarks). The short-distance cross sections $d \hat\sigma$ are
essentially inclusive partonic cross sections for producing the single
parton $i$, convolved with parton distributions if the colliding
particles $A$ and $B$ are hadrons. They have perturbative expansions in
powers of $\alpha_s(p_T)$. The fragmentation function $D_{i\to H}(z)$ is
the nonperturbative probability distribution in the longitudinal
momentum fraction $z$ of the hadron $H$ relative to the parton $i$. The
leading power (LP) factorization formula in Eq.~(\ref{1-ple-frag}) was
proven for $e^+ e^-$ annihilation by Collins and Soper
\cite{Collins:1981uw}. For a light hadron $H$, the corrections are of
order $\Lambda_{\rm QCD}^2/p_T^2$. The proof does not seem to have been
extended to hadron collisions, except in the case of heavy quarkonium,
for which a proof has been sketched recently by Nayak, Qiu, and Sterman
\cite{Nayak:2005rt}. In this case, the leading corrections are of order
$m_Q^2/p_T^2$.

The LP fragmentation formula in Eq.~(\ref{1-ple-frag}) has been used to
describe quarkonium production at large $p_T$. The fragmentation
functions for a heavy quarkonium can be calculated in perturbative QCD,
up to nonperturbative multiplicative constants
\cite{Braaten:1993rw,Braaten:1993mp}. The fragmentation functions have
been calculated in the NRQCD factorization framework to leading order in
$\alpha_s$ for all the phenomenologically relevant channels and to
next-to-leading order (NLO) for the color-octet $^3S_1$ channel
\cite{Braaten:2000pc}. In some channels, the leading power of $p_T$
described by parton fragmentation does not appear in the NRQCD
factorization formula until NLO or even N$^2$LO in $\alpha_s$.  For
these channels, the LP fragmentation formula in Eq.~(\ref{1-ple-frag})
can be used to calculate the leading power of $p_T$ with much less
effort than a fixed-order calculation. Furthermore, the evolution
equations for the single-parton fragmentation functions can be used to
sum large logarithms of $p_T^2/m_Q^2$ to all orders in $\alpha_s$.
Unfortunately, the usefulness of the LP fragmentation formula in
Eq.~(\ref{1-ple-frag}) for quarkonium has proved to be limited.
Explicit calculations have revealed that, in some channels, it does
not give the largest contribution until $p_T$ is almost an order of
magnitude larger than the quarkonium mass
\cite{Chang:1994aw,Chang:1996jt}. In these channels, the corrections
of relative order $m_Q^2/p_T^2$ apparently have large coefficients.

An important recent theoretical development is the extension of the
factorization formula in Eq.~(\ref{1-ple-frag}) to the first subleading
power of $p_T^2$. Kang, Qiu, and Sterman proved that the
contributions of relative order $m_Q^2/p_T^2$ can be written as
a sum of of $Q\bar Q$ production cross sections convolved with
{\it double-parton fragmentation functions}
\cite{Kang:2011zza,Kang:2011mg}:
\begin{equation}
\sum_n d\hat\sigma[A+B\to (Q\bar Q)_n +X]\otimes D[(Q\bar Q)_n \to H].
\label{2-ple-frag}
\end{equation}
The sum over $n$ extends over the color (singlet or octet) and Lorentz
structures (vector or axial vector) of the energetic $Q \bar Q$ pair.
The short-distance cross sections $d \hat\sigma$ are essentially
inclusive partonic cross sections for producing an energetic $Q \bar Q$
pair, convolved with parton distributions if the colliding particles $A$
and $B$ are hadrons. They have perturbative expansions in powers of
$\alpha_s(p_T)$. The double-parton fragmentation functions 
$D_{(Q\bar Q)_n \to H}(z, \zeta,\zeta')$ are nonperturbative probability
distributions in the longitudinal momentum fraction $z$ of the
quarkonium $H$ relative to the $Q \bar Q$ pair. 
They also depend on the relative longitudinal
momentum fractions $\zeta$ and $\zeta'$ of the $Q$ and the $\bar
Q$. The evolution equations for the double-parton fragmentation
functions can be used to sum large logarithms of $p_T^2/m_Q^2$ to all
orders in $\alpha_s$. The NLP fragmentation
formula is obtained by adding Eq.~(\ref{2-ple-frag}) to Eq.~(\ref
{1-ple-frag}). The corrections are of order $\Lambda_{\rm QCD}^2/p_T^2$
and $m_Q^4/p_T^4$. This factorization formula has also been derived by
making use of soft collinear effective theory
\cite{Fleming:2012wy,Fleming:2013qu}.

The NLP fragmentation approach by itself lacks the predictive power of
NRQCD factorization, because the fragmentation functions $D[i\to H]$ and
$D[(Q\bar Q)_n \to H]$ are nonperturbative functions of the momentum
fractions that must be determined phenomenologically. The NLP
fragmentation approach can be given predictive power through the NRQCD
factorization conjecture, which implies that a fragmentation function
can be written as a sum of products of short-distance functions times
NRQCD matrix elements, in analogy to Eq.~(\ref{NRQCD-fact}).  If one
truncates the expansion in the relative velocity $v$, as is described
in Sec.~\ref{sec:nrqcd-fact}, then the nonperturbative factors can be
reduced to a few constants for each quarkonium spin multiplet. The NLP
fragmentation cross section, expressed in terms of the NRQCD matrix
elements, should account for the first two terms in the expansion of the
NRQCD factorization cross section in powers of $m_Q^2/p_T^2$.

\subsubsection{Status of Proofs of NRQCD Factorization}

A diagrammatic proof of NRQCD factorization for a given hard-scattering
process consists of a demonstration that diagrams in each order in
$\alpha_s$ can be reorganized so that (1) all soft singularities cancel
or can be absorbed into NRQCD matrix elements and (2) all collinear
singularities and spectator interactions can be absorbed into parton
distributions of incoming hadrons. This reorganization of
low-virtuality singularities is essential in order to define $Q \bar
Q$ production cross sections that are free of low-momentum contributions
and, hence, calculable in perturbation theory.

In the NLP fragmentation approach, the factorization of the production
cross section into the form of Eqs.~(\ref{1-ple-frag}) and
(\ref{2-ple-frag}), up to corrections that are suppressed by factors of
$\Lambda_{\rm QCD}^2/p_T^2$ or $m_Q^4/p_T^4$, has already been proven
\cite{Kang:2011zza,Kang:2011mg,Fleming:2012wy}. It remains to prove
that the fragmentation functions satisfy the NRQCD factorization
conjecture. This step has been demonstrated for gluon fragmentation
functions through next-to-next-to-leading order (NNLO) in $\alpha_s$
\cite{Nayak:2005rw,Nayak:2005rt,Nayak:2006fm}. However, a proof to all
orders in $\alpha_s$ is essential, because potential violations of
factorization involve soft gluons, for which $\alpha_s$ is not a good
expansion parameter. 

In the absence of a rigorous proof of NRQCD factorization, we must rely
on experiment in order to decide whether NRQCD factorization is a valid
model of quarkonium production. Standard methods for proving
factorization of soft contributions are valid only up to corrections
that are suppressed as $1/p_T^4$. In particular, the NLP fragmentation
approach suggests that the NRQCD factorization formula in
Eq.~(\ref{NRQCD-fact}) is likely to hold only up to corrections that
are suppressed by a factor $m_Q^4/p_T^4$ or a factor
$\Lambda_{\rm QCD}^2/p_T^2$. Therefore, it is very important to test the
predictions of NRQCD factorization with high precision at the highest
accessible values of $p_T$.

\subsubsection{$k_T$ Factorization}

The $k_T$-factorization approach is an alternative to standard collinear
factorization in which one writes cross sections in terms of parton
distributions that depend on the transverse momenta of the partons, as
well as on their longitudinal momentum fractions. The
$k_T$-factorization approach contains some contributions at leading
order in $\alpha_s$ that would appear only in higher orders in collinear
factorization. This property may be useful in certain kinematic
situations in which large corrections can appear in higher orders in
calculations in collinear factorization. The predictions of the
$k_T$-factorization approach agree well with many existing measurements
of quarkonium production cross sections and polarizations. However,
because the $k_T$-dependent parton distributions are known
phenomenologically with much less precision than the collinear parton
distributions, $k_T$-factorization predictions can contain large
uncertainties that may not yet be accurately quantified. Moreover, one
must be cognizant of the fact that there could be hidden biases in
the predictions that stem from the choices of parametrizations of the
$k_T$-dependent parton distributions. Furthermore, in practice, the
$k_T$-dependent parton distributions model large-$k_T$ behavior that
would be calculated from first principles in collinear factorization.
                     
In the applications of $k_T$ factorization to quarkonium production, it
is always assumed that production occurs only through color-singlet
$Q\bar Q$ channels. Therefore, the $k_T$-factorization approach
contains uncanceled infrared divergences, which appear in higher orders
in $v$, for example, in calculations of $P$-wave quarkonium production
at the leading non-trivial order in $v$.

\subsubsection{Large Higher-Order Perturbative Corrections}
\label{sec:HOCor}

The short-distance coefficients $d \sigma_n$ in the 
NRQCD factorization formula in Eq.~(\ref{NRQCD-fact}) 
are partonic cross sections for creating a $Q\bar Q$ pair
that can be calculated as perturbation series in $\alpha_s$.
For most of the important production processes, 
they have been calculated to next-to-leading order (NLO) in $\alpha_s$.
In some cases, the NLO corrections are uncomfortably large.
Understanding the origin of the large radiative corrections 
can lead to new insights into the production mechanisms.

In inclusive quarkonium production at large $p_T$, corrections of NLO in
$\alpha_s$ are surprisingly large  in some channels --- sometimes
exceeding the leading-order (LO) contribution by an order of magnitude.
This situation has raised questions about the convergence of the
perturbation expansion. The large higher-order corrections may be
understood as arising from a combination of two effects. The first 
effect is that higher powers of $\alpha_s$ can be offset by a less rapid
fall-off with $p_T$. For example, for production of the $J/\psi$ through
the color-singlet $S$-wave channel, the contribution to $d
\sigma/dp_T^2$ at leading order in $\alpha_s$ goes as $\alpha_s^3
m_c^4/p_T^8$, and so it is strongly suppressed at large $p_T$. The NLO
contribution goes as $\alpha_s^4 m_c^2/p_T^6$, so it is less strongly
suppressed.  The NNLO contribution goes as $\alpha_s^5/p_T^4$, owing to
contributions in which a gluon fragments into the $J/\psi$. Higher
orders exhibit this same leading-power behavior, $1/p_T^4$, but
are suppressed by additional powers of $\alpha_s$. The second effect is
that the contributions from $Q \bar Q$ fragmentation into quarkonium can
enter with large coefficients.  Gluon fragmentation contributions scale
as the leading power $1/p_T^4$, while the  $Q \bar Q$ fragmentation
contributions scale as $m_Q^2/p_T^6$, and so gluon fragmentation will
dominate at sufficiently large $p_T$. However, as we have mentioned,
there are channels in which  both contributions enter at the same order
in $\alpha_s$, but the value of $p_T$ at which gluon fragmentation
begins to dominate is almost an order of magnitude larger than the
quarkonium mass ~\cite{Chang:1994aw,Chang:1996jt}.

The NLP fragmentation approach has the potential to increase
significantly the accuracy of predictions for inclusive quarkonium
production at large $p_T$. In many color and angular-momentum channels,
the accuracy of the existing NLO calculations is at best leading order
at large $p_T$. At very large $p_T$, the accuracy is at best leading
order because single-parton fragmentation contributions that behave
as $1/p_T^4$ enter first at NLO in a strict expansion in $\alpha_s$.
At intermediate $p_T$, the accuracy is at best leading order because
$Q \bar Q$ fragmentation contributions that may have large
coefficients and that behave as $m_Q^2/p_T^6$ enter first at NLO.
In fact, the accuracy in these channels is not even leading order,
because there are large logarithms of $p_T/m_Q$ at higher orders in
$\alpha_s$. Within NRQCD factorization, a complete calculation to NNLO
in $\alpha_s$ would be prohibitively difficult and would only decrease
the relative error to $\alpha_s \log(p_T/m_Q)$ at large $p_T$. Within
the fragmentation approach, the relative error could be decreased to
order $\alpha_s^2$ at large $p_T$ by calculating the fragmentation
functions to NLO in $\alpha_s$ and using the evolution equations for the
fragmentation functions to sum the logarithms of $p_T/m_Q$. Such a
calculation is feasible, and it would give higher accuracy at large
$p_T$ than a full calculation to NNLO in $\alpha_s$.

\subsubsection{Exclusive quarkonium production}
 
A collision that is initiated by an electron and a positron or by two
photons can produce exclusive two-quarkonium final states. The exclusive
production rate can be calculated, at sufficiently large CM momentum
$p_T$, by making use of a simplified version of the NRQCD factorization
formula in Eq.~(\ref{NRQCD-fact}) in which the color-octet NRQCD matrix
elements are omitted. The corrections to this formula should be
suppressed by at least $v^4$. In this simplified factorization formula,
the inclusive partonic production cross sections are, of course,
replaced by the appropriate exclusive cross sections.

For exclusive $e^+e^-$ production of two quarkonium states and for
exclusive production of a quarkonium and a light meson in $B$-meson
decays, factorization theorems have been established for processes that
proceed at leading order without a helicity flip of the heavy-quark
\cite{Bodwin:2010fi}. For exclusive processes that proceed through a
helicity flip at leading order, such as $e^+e^-\to J/\psi +\eta_c$,
factorization has not been proven because there are complications from
the ``endpoint regions'' of momentum space, in which a gluon carries
away most of the longitudinal momentum of a spectator quark
\cite{Bodwin:2013ys}. It is possible that NRQCD factorization still
holds in these cases because the endpoint regions cannot have a
virtuality lower than $m_Q$.

Exclusive quarkonium production cross sections have also been calculated
within the light-cone formalism for exclusive hard-scattering processes
\cite{Chernyak:1977fk,Efremov:1979qk,Lepage:1979za,Lepage:1979zb,Lepage:1980fj,Chernyak:1980dk}.
Examples of such calculations can be found in
Refs.~\cite{Bondar:2004sv,Braguta:2008tg} and references therein.
Calculations in the light-cone approach neglect parton transverse
momenta in comparison with longitudinal momenta and are, therefore,
sometimes simpler than calculations in the NRQCD factorization approach
\cite{Jia:2008ep}. In first approximation, the light-cone approach takes
into account certain contributions that are of higher order in $v$ in
the NRQCD factorization approach. However, existing calculations in the
light-cone approach rely on the use of model light-cone quarkonium
distributions. In some calculations, the small-momentum behavior of the
model light-cone distributions has been constrained by making use of
information about the quarkonium wave function. However, the
large-momentum tails of the model light-cone distributions may not be
consistent with the large-momentum behavior of QCD \cite{Bodwin:2006dm}.
An important limitation of the light-cone approach is that, because it
neglects the transverse momentum of partons, it does not reproduce the
endpoint regions of QCD properly \cite{Jia:2010fw,Bodwin:2013ys}. The
endpoint regions give contributions that are enhanced by logarithms of
$p_T^2/m_Q^2$ in helicity-flip processes, such as $e^+e^-\to J/\psi
+\eta_c$.

\subsection{Issues That Should Be Addressed in Future Work}

NLO corrections in the NRQCD factorization framework have been computed
for many quarkonium production processes.  NLO computations for
inclusive production include $J/\psi$, $\psi(2S)$, $\chi_{cJ}$ and
$\Upsilon(nS)$ production cross sections and polarizations at the
Tevatron and the LHC,  $J/\psi$ and $\psi(2S)$ production cross sections
at RHIC, $J/\psi$ photoproduction cross sections and polarization at
HERA, the $J/\psi+\eta_c$ production cross section, and the $J/\psi+X$
and $J/\psi+X(\hbox{non-$c\bar c$})$ production cross sections in
$e^+e^-$ annihilation at the $B$ factories. NLO computations for
exclusive production include the $e^+e^-\to J/\psi+\eta_c$ and
$e^+e^-\to J/\psi+\chi_{cJ}$ double-quarkonium production cross
sections. Generally, data and the NLO predictions of NRQCD
factorization for quarkonium production agree, within
errors.\footnote{See the global fit of NRQCD matrix elements of
Butensch\"on and Kniehl \cite{Butenschoen:2011yh,Butenschoen:2012qh} and
Ref.~\cite{Brambilla:2010cs} for some further details.} There are three
significant exceptions. These are discussed below, along with some
additional issues that should be addressed in future experimental and
theoretical work.

\subsubsection{$e^+e^-\to J/\psi + X({\rm non-}c \bar c)$}

Inclusive charmonium production in $e^+e^-$ collisions at the $B$
factories has shown some discrepancies with theoretical predictions. The
inclusive cross section for $J/\psi$ production can be split into the
cross section to produce $J/\psi + c\bar c +X$, in which there are charm
mesons accompanying the $J/\psi$, and the cross section to produce
$J/\psi + X({\rm non-}c \bar c)$, in which there are only light hadrons
accompanying the $J/\psi$. The measured cross section for $J/\psi +
X({\rm non-}c \bar c)$ production at Belle \cite{Pakhlov:2009nj} is
about a factor two lower than NRQCD factorization predictions
\cite{Zhang:2009ym,Butenschoen:2011yh}. However, for several
reasons, the apparent conflict between the Belle measurement and the
theoretical predictions may not be significant. First, the typical
$J/\psi$ CM momentum $p_T$ for this process is quite low --- less than
3~GeV --- suggesting that corrections to factorization of order
$m_c^4/p_T^4$ may not be under control. Second, the most recent Belle
measurements \cite{Pakhlov:2009nj} imply a value for the inclusive
$J/\psi$ production cross section that is about a factor of two smaller
than the value that was measured by the BaBar Collaboration
\cite{Aubert:2001pd}. Independent measurements of the cross sections to
produce $J/\psi + X$ and $J/\psi + X({\rm non-}c \bar c)$ would
therefore be valuable. Third, the Belle measurement of cross sections
for $J/\psi$ plus light hadrons includes only events with greater than
four charged tracks, and the corrections owing to events with four or
fewer charged tracks are not known. A determination of the effect of
events with four or fewer charged tracks is necessary in order to
make a direct comparison between theory and experiment.

\subsubsection{$\gamma\gamma \to J/\psi + X$}

The cross section for inclusive $J/\psi$ production in $\gamma\gamma$
scattering that was measured by DELPHI at LEP~II \cite{Abdallah:2003du}
lies above the NRQCD factorization prediction that is based on the
global fits of NRQCD matrix elements \cite{Butenschoen:2011yh} by more
than an order of magnitude. However, the experimental error bars are
very large, and the discrepancies with the prediction all lie at values
of $p_T$ less than $2.7$~GeV, where corrections to factorization of
order $m_c^4/p_T^4$ may not be under control. Clearly, measurements of
greater precision and at higher values of $p_T$ are needed in order to
make a meaningful comparison with the prediction of NRQCD factorization.

\subsubsection{$J/\psi$ and $\Upsilon$ Polarization}

The $J/\psi$ polarization that is observed in CDF Run~II
\cite{Abulencia:2007us} seems to be incompatible with the combined
constraints of the measured $d\sigma/dp_T$ from CDF Run~II
\cite{Abulencia:2007us} and the measured $d\sigma/dp_T$ from HERA
\cite{Adloff:2002ex,Aaron:2010gz}. The CDF Run~II measurement
indicates that the $J/\psi$ is slightly longitudinally polarized in the
helicity frame. NRQCD predictions for polarization of the $J/\psi$ vary
dramatically, depending on the data that are used to determine the NRQCD
matrix elements. A prediction of strong transverse polarization
\cite{Butenschoen:2012px} arises when one includes in the fits of the
NRQCD matrix elements both the CDF Run~II data and the HERA data for
$d\sigma/dp_T$ down to a $p_T$ of $3$~GeV. Inclusion of the low-$p_T$
HERA data is crucial in obtaining strong transverse polarization. At
such low values of $p_T$, one might doubt that corrections to
factorization of order $m_c^4/p_T^4$ are under control. A prediction of
moderate transverse polarization \cite{Gong:2012ug} arises when one
includes data for $d\sigma/dp_T$ with $p_T$ greater than $7$~GeV from
both CDF Run~II and LHCb and uses NRQCD factorization predictions to
correct for feeddown from the $\psi(2S)$ and the $\chi_{cJ}$ states. A
prediction of near-zero transverse polarization \cite{Chao:2012iv}
arises when one includes in the fits of the NRQCD matrix elements both
the CDF Run~II data for $d\sigma/dp_T$ with $p_T$ greater than $7$~GeV
and the CDF Run~II polarization measurement. It is very important to
resolve the ambiguities in the fits of the NRQCD matrix elements.
Further measurements of $J/\psi$ production in new processes, at
different values of $\sqrt{s}$ and rapidity, and in $ep$ collisions at
larger values of $p_T$ would all help to resolve these ambiguities. It
is worth noting that the CDF Run~I \cite{Affolder:2000nn} and Run~II
\cite{Abulencia:2007us} $J/\psi$ polarization measurements are
incompatible, although the CDF collaboration states
\cite{Abulencia:2007us} that the Run~I measurement supersedes the Run~I
measurement. Recent results from $J/\psi$ polarization measurements at
LHCb \cite{Aaij:2013hsa} and from $J/\psi$ and $\psi(2S)$ polarization
measurements at CMS \cite{Chatrchyan:2013cla} show no evidence for large
transverse polarization in the helicity frame out to $p_T=57$~GeV.

CDF Run~II polarization measurements for all three $\Upsilon$(nS)
states have been analyzed in several different polarization frames.
A frame-invariant relation has been used to check the consistency of
these measurements~\cite{CDF:2011ag}. These Run~II results agree with
the Run~I CDF results in the helicity frame and extend those results to
larger values of $p_T$. The Run~II \DZero\ result for the $\Upsilon(1S)$ in
the helicity frame disagrees with both CDF measurements, but has much
less precision than the new CDF result. A recent CMS measurement of the
$\Upsilon$(nS) polarizations in pp collisions at 7~TeV shows behavior
that is similar to that of the CDF Tevatron results in multiple
reference frames and in a frame-independent analysis
\cite{Chatrchyan:2012woa}.  The trends of the $\Upsilon$ measurements as
functions of $p_T$, in both $pp$ and $p\bar p$ production, agree with
the trends in the charmonium system: there are no large polarization
effects at high $p_T$. A new NLO prediction for the polarizations of the
$\Upsilon(nS)$ states at the LHC \cite{Gong:2013qka} is in agreement
with data from the CMS collaboration \cite{Chatrchyan:2012woa} for the
$\Upsilon(1S)$ and $\Upsilon(2S)$ states, but not for the $\Upsilon(3S)$
state for $p_T >$ 30 GeV. The agreement of the new predictions with CDF
Run~II data is not quite as good. Clearly, it will be important to
address the issue of feeddown from higher-mass $b \bar b$ states in future
experimental measurements of $\Upsilon(nS)$ polarizations.

\subsubsection{Need for Resummation of Logs of $p_T^2/m_c^2$}

LHC data for $J/\psi$ production, particularly data at large values of
$p_T$, show a slight shape discrepancy in comparison with NRQCD
predictions --- especially the predictions of Butensch\"on and Kniehl
\cite{Butenschoen:2011yh}, which are based on global fits of the NRQCD
matrix elements that include data with $p_T$ as low as $3$~GeV. This
shape discrepancy may be an indication that resummation of logarithms of
$p_T^2/m_c^2$ is needed in the NRQCD factorization predictions at large
$p_T$. The NLP fragmentation approach provides a framework with which
to carry out this resummation at both the leading power and the first
subleading power in $m_Q^2/p_T^2$.

\subsubsection{The Issue of Feeddown}

Comparisons between theory and experiment are complicated by feeddown
from heavier quarkonium states.  The prompt production rates for
$J/\psi$ include feeddown from the $\psi(2S)$ and $\chi_{cJ}(1P)$ states
and the prompt production rates for the $\Upsilon(1S)$ include
feeddown from the $\Upsilon(2S)$, $\Upsilon(3S)$, $\chi_{bJ}(1P)$, and
$\chi_{bJ}(2P)$ states. Theoretical predictions, in contrast, are
typically for direct production rates of the $J/\psi$ and $\Upsilon(1S)$
states. Theoretical predictions have been made that include the effects
of feeddown. However, it should be kept in mind that they are obtained
by adding the predictions for several direct production rates, which may
not be equally reliable. In the case of unpolarized cross sections, the
feeddown contributions are typically on the order of about
$30$--$40\%$, which is not significant in comparison with the theoretical
uncertainties. However, the feeddown contributions could have an
important effect on the polarizations. Given these considerations, it
would be very useful for experiments to separate the feeddown
contributions from the direct-production contributions, especially for
polarization studies.

Measurements of $\chi_{cJ}$ and $\chi_{bJ}$ production rates and
polarizations would provide additional tests of NRQCD factorization. For
the $\chi$ states, the theory is more constrained than for the $\psi$
and $\Upsilon$ states because there are two, rather than four, NRQCD
matrix elements that enter at the leading non-trivial order in $v$.
Consequently, it might be possible to make more stringent tests of NRQCD
factorization for the $\chi_{cJ}$ and $\chi_{bJ}$ states than for the
$S$-wave quarkonium states.

\subsubsection{Universality of NRQCD matrix elements}

Universality of the NRQCD matrix elements is a crucial prediction of the
NRQCD factorization approach. It should be tested in as wide a range of
production processes as possible. In the case of spin-triplet S-wave
states, measurements of associated production of quarkonia or
measurements of photoproduction or leptoproduction processes, in
addition to hadroproduction processes, may be essential in order to pin
down the values of the three most important color-octet NRQCD matrix
elements. Those values are needed in order to make firm predictions of
cross sections and, especially, polarizations.

\subsubsection{Large Logarithms in Exclusive Double-Quarkonium Production}

NRQCD factorization predictions for exclusive double-quarkonium
production in $e^+e^-$ annihilation generally agree with experimental
measurements, within rather large experimental and theoretical
uncertainties. For many years, a large discrepancy existed between
theoretical predictions and experimental measurements of the cross
section for $e^+e^-\to J/\psi+\eta_c$. This discrepancy was resolved by
a shift in the measured cross section and by the incorporation of
corrections of higher order in $v$ and $\alpha_s$ into the theoretical
predictions. The corrections of next-to-leading order (NLO) in $v^2$ and
the corrections of NLO in $\alpha_s$ each increase the cross section by
about a factor of two. The large size of the corrections of NLO in $v^2$
is not believed to be a problem for the $v$ expansion because these
corrections arise from three different sources, each of which seems to
have a well-behaved $v$ expansion \cite{Bodwin:2006ke,Bodwin:2007ga}.
The large size of the corrections of NLO in $\alpha_s$ is more
problematic.  It is a consequence of large double and single
logarithms of $p_T^2/m_c^2$ \cite{Jia:2010fw}. It has been shown that
the double logarithms arise from the endpoint region
\cite{Bodwin:2013ys}, in which the hard scattering transfers almost all
of the momentum of a spectator parton to an active parton. However, it
is not known how to resum the double (or single) endpoint logarithms to
all orders in perturbation theory. Such a resummation would allow one to
have greater confidence in the reliability of the theoretical
predictions.

\subsection{Opportunities at the Frontiers of High-Energy Physics}

In this section, we describe the opportunities in the near- and
long-term future for adding to our understanding of quarkonium
production. We describe the prospects for progress in this arena at
the Large Hadron collider, Belle-II, the LHC upgrade, and future
high-energy $e^+ e^-$ and $e p$ colliders.

\subsubsection{Large Hadron Collider}

The extension of the energy frontier to 13 TeV in Run 2 of the LHC will
offer the prospects of an extended $p_T$ reach for quarkonium studies. 
The experiments are refining their trigger strategies to cope with the
higher luminosity and  the  larger cross section.  It is important that
quarkonium studies should not be ignored in the emphasis on the search
for new physics. The importance of quarkonium decay modes of the Higgs
boson as probes of Higgs coupling constants was mentioned in
Sec.~\ref{sec:probes-new-physics}.  The exploration of the high-$p_T$
aspects of quarkonium production and polarization that are outlined here
will require large integrated luminosity.  In this regard, it is
important that the experiments ensure that quarkonium triggers retain
high efficiency for dimuon decays of the $J/\psi$ and the $\Upsilon(1S)$
at large $p_T$.

The current benchmark theoretical model for quarkonium production, NRQCD
factorization, has not been proven by theoretical means to be a
consequence of QCD. In the absence of further progress in establishing
NRQCD factorization theoretically, we must rely on experiment to decide
whether it is a valid model of quarkonium production. The NLP
fragmentation approach suggests that NRQCD factorization is likely to
hold only for $p_T$ much greater than the quarkonium mass. 
Therefore, it is very important to measure quarkonium production rates
and polarizations with high statistics at the highest possible values of
$p_T$. Measurements of production cross sections differential in both
rapidity and $p_T$ could provide additional important tests, as could
measurements of the production rates and polarizations of $\chi_{cJ}$
and $\chi_{bJ}$ states. In making measurements of prompt quarkonium
production, it is very important to separate feeddown contributions from
the direct-production rates. Measurements of new production processes,
such as the associated production of a quarkonium state and a $W$ or $Z$
boson, would also yield important tests of theoretical models.

The LHCb detector has demonstrated its ability to measure the
properties of quarkonium states in a variety of decay modes.  In
the case of $B_c$ studies, the luminosity limitation may restrict 
the range of final states that can be addressed, but the experiment has
already shown its ability to make good determinations of relative
branching ratios and should have enough data in Run~2 to explore
the physics implications of a quarkonium state that has net flavor
content.  It is not clear whether the experiment has a chance to
observe any of the $B_c$ excited states.

\subsubsection{Belle-II at SuperKEKB}

Double-charmonium production processes have been the focus of
interesting measurements at the $B$ factories that were not anticipated
in the designs of these machines.  At Belle, the cross sections for
producing $J/\psi + \eta_c(1S)$, $J/\psi + \chi_{c0}(1P)$, and $J/\psi +
\eta_c(2S)$ have all been measured. The large increase in luminosity at
Belle~II should make it possible to resolve processes in which the
$J/\psi$ recoils against other onia. In particular, it should be
possible to measure the cross section for  producing $J/\psi + J/\psi$,
which proceeds through the annihilation of $e^+ e^-$ into two photons
\cite{Bodwin:2002kk}. It may also be possible to observe processes in
which a $\chi_{cJ}(1P)$ or  an  $\eta_c(1S)$ recoils against an onium.

As we have mentioned in Sec.~\ref{sec:HOCor}, the theory of
exclusive production of double-quarkonium states is plagued by large
logarithms of $p_T^2/m_Q^2$, where $p_T$ is the CM momentum of the
quarkonium. These logarithms appear in channels for which the production
amplitude  is helicity suppressed  at leading order in $\alpha_s$.
Such channels include $J/\psi + \eta_c$ and, for some combinations of
helicities, $J/\psi + \chi_{cJ}$.  Work is in progress with the goal of
understanding the origins of such large logarithms and resumming them to
all orders in perturbation theory. If the perturbation series can be
brought under control through such a resummation, then the resummation
could be tested by measuring the exclusive production of
double-charmonium states at Belle-II.  Measurements of rates into
specific helicity states would provide particularly stringent tests of
the theory.

If the center-of-mass energy at Belle-II can be increased above the
$B_c^+ B_c^-$ threshold at 12.5~GeV, that would allow measurements
of decays of the $B_c$ that are hopeless at a hadron collider. It might
even allow the observation of some of the excited states in the $B_c$
spectrum.

\subsubsection{LHC Upgrade}
 
It is likely that many of the tests of quarkonium-production theory that
we have outlined will be accomplished in the upcoming runs of the LHC.
The higher energy and luminosity of an upgraded LHC might however be of
considerable importance for testing the theory of the production of
$b\bar b$ quarkonium states, as the criterion that $p_T$ be much larger
than the bottomonium mass may not be easy to satisfy in precision
measurements. Additional precision in measurements of charmonium
production at higher values of $p_T$ could also drive the theory to a
new level of precision beyond NLO in $\alpha_s$ and beyond the leading
non-trivial order in $v$.

\subsubsection{Future $e^+e^-$ Collider}

One important application of a future $e^+e^-$ collider will be to serve
as a Higgs factory.  This will make possible the measurements of Higgs
couplings to the SM particles.  As we have mentioned 
in Sec.~\ref{sec:probes-new-physics}, rare Higgs decays involving
quarkonia offer unique probes of some couplings.
In particular, the decay $H\to J/\psi + \gamma$ seems to provide 
the only realistic probe of the $H c \bar c$ coupling.

Measurements of inclusive quarkonium production in $\gamma \gamma$
collisions at a high-energy $e^+ e^-$ collider could provide an important
text of NRQCD factorization.  In particular, measurements of inclusive
$J/\psi$ production at higher values of $p_T$ than were accessible at
LEP could resolve the tension between global fits of NRQCD matrix
elements and the LEP measurements.

\subsubsection{Future $ep$ Collider}

The large range of theoretical predictions for quarkonium polarizations
in hadroproduction is a consequence of the fact that the color-octet
NRQCD matrix elements are poorly determined at present by the high-$p_T$
hadroproduction data. Such ambiguities could likely be largely resolved
by measuring quarkonium production cross sections with high precision at
$p_T> 10$~GeV at a future $ep$ collider. Measurements of quarkonium
polarization at a high-energy $ep$ collider would provide further
stringent tests of the theory. 

An electron-ion collider has emerged as one of the top priorities of the
U.S.\ nuclear-physics community. An electron-ion collider at CERN has
also been proposed.

\appendix

\begin{acknowledgments}

We thank Vaia Papadimitriou for helpful comments. The work of TKP was
supported in part by the U.S. National Science Foundation under Grant
No. PHY-1205843. The work of SLO was supported by the National Research
Foundation of Korea under Grant No. 2011-0029457 and WCU Grant
No. R32-10155. The work of EB was supported in part by the U.S.
Department of Energy under grant No. DE-FG02-91-ER40690. The work of JR
was supported in part by the U.S. Department of Energy under Grant No.
DE - SC0010118TDD. The work of EE was supported by Fermilab, which is
operated by Fermi Research Alliance, LLC, for the U.S. Department of
Energy under Contract No. DE-AC02-07CH11359. The work of GTB was
supported by the U.S. Department of Energy, Division of High Energy
Physics, under Contract No. DE-AC02-06CH11357. The submitted manuscript
has been created in part by UChicago Argonne, LLC, operator of Argonne
National Laboratory (Argonne). Argonne, a U.S. Department of Energy
Office of Science laboratory, is operated under Contract No.
DE-AC02-06CH11357. The U.S. Government retains for itself, and others
acting on its behalf, a paid-up nonexclusive, irrevocable worldwide
license in said article to reproduce, prepare derivative works,
distribute copies to the public, and perform publicly and display
publicly, by or on behalf of the Government.

\end{acknowledgments}

\end{document}